\documentclass[acmsmall,screen]{acmart}

\usepackage{xspace}
\usepackage{multirow}

\usepackage{mdframed}
\usepackage{framed}
\usepackage{xcolor}
\usepackage[ruled, lined, linesnumbered, commentsnumbered, longend]{algorithm2e}
\usepackage{tablefootnote}

\usepackage{graphicx}
\usepackage{booktabs}

\SetCommentSty{mycommfont}

\mdfdefinestyle{MyShadeQuoteStyle}{%
    leftmargin=0pt,
    rightmargin=0pt,
    backgroundcolor=gray!25,
    linewidth=0pt,
    skipbelow=\topskip,
    skipabove=\topskip
}

{\begin{snugshade}\begin{quote}}
{\hfill\end{quote}\end{snugshade}}
\definecolor{shadecolor}{rgb}{0.9,0.9,0.9}
\definecolor{darkyellow}{rgb}{0.5, 0.4, 0.0}

\usepackage{tcolorbox}
\newcommand{\rqbox}[1]{\begin{tcolorbox}[left=2pt,right=2pt,top=2pt,bottom=2pt,colback=gray!4,colframe=gray!40!black,before skip=6pt,after skip=7pt]#1\end{tcolorbox}}

\newcommand{\rqone}{\textbf{RQ1: How effective is our approach \ourTool?}}

\newcommand{\rqtwo}{\textbf{RQ2: What is the contribution of individual components to \ourTool?}}

\newcommand{\rqthree}{\textbf{RQ3: How do parameters affect the effectiveness of \ourTool?}}

\newcommand{\rqfour}{\textbf{RQ4: Does few-shot learning improve \ourTool?}}

\newcommand{\ourTool}{\textit{ZS4C}\xspace}
\newcommand{\ourToolGPTThree}{\textit{ZS4C$_{GPT-3.5}$}\xspace}
\newcommand{\ourToolGPTFour}{\textit{ZS4C$_{GPT-4}$}\xspace}

\newcommand{\basicprompt}{\textit{BasePrompt}\xspace}
\newcommand{\basicpromptGPTThree}{\textit{BasePrompt$_{GPT-3.5}$}\xspace}
\newcommand{\basicpromptGPTFour}{\textit{BasePrompt$_{GPT-4}$}\xspace}

\newcommand{\snr}{SnR\xspace}

\newcommand{\inference}{\textit{ImportStInfer}\xspace}
\newcommand{\compliationEn}{\textit{ConversationFixing}\xspace}
\newcommand{\inferenceWOSC}{\textit{ImportStInfer$_{noSC}$}\xspace}

\newcommand{\inferenceFullN}{Import Statement Inference\xspace}
\newcommand{\compilationEnFullN}{Conversational Error Fixing\xspace}

\usepackage{listings}

\usepackage{xcolor}

\usepackage{array}

\usepackage{caption}
\usepackage{subcaption}

\definecolor{codegreen}{rgb}{0,0.6,0}
\definecolor{codegray}{rgb}{0.5,0.5,0.5}
\definecolor{codepurple}{rgb}{0.58,0,0.82}
\definecolor{backcolour}{rgb}{0.95,0.95,0.92}

\lstdefinestyle{javaStyle}{
    basicstyle=\ttfamily\footnotesize,
    keywordstyle=\color{blue},
    commentstyle=\color{gray},
    stringstyle=\color{red},
    showstringspaces=false,
    language=Java,
    frame=single,
    numbers=none,
    numberstyle=\tiny,
    stepnumber=1,
    numbersep=5pt,
    aboveskip=10pt,
    belowskip=10pt,
    escapeinside={(*@}{@*)},
}

\newtcbox{\errorbox}[1][red!20]{
    on line,
    arc=0pt,
    outer arc=0pt,
    colback=#1,
    colframe=#1,
    boxrule=0pt,
    boxsep=0pt,
    left=1pt,
    right=1pt,
    top=1pt,
    bottom=1pt,
}

\newtcbox{\fixbox}[1][green!20]{
    on line,
    arc=0pt,
    outer arc=0pt,
    colback=#1,
    colframe=#1,
    boxrule=0pt,
    boxsep=0pt,
    left=1pt,
    right=1pt,
    top=1pt,
    bottom=1pt,
}

\lstdefinestyle{styleformodificationdate}{
    backgroundcolor=\color{white}, 
    basicstyle=\ttfamily,          
    numbers=none,                  
    frame=none                     
}

\lstdefinestyle{mystyle}{
  backgroundcolor=\color{backcolour}, commentstyle=\color{codegreen},
  keywordstyle=\color{magenta},
  numberstyle=\tiny\color{codegray},
  stringstyle=\color{codepurple},
  basicstyle=\ttfamily\footnotesize,
  breakatwhitespace=false,         
  breaklines=true,                 
  captionpos=b,                    
  keepspaces=true,                 
  numbers=left,                    
  numbersep=5pt,                  
  showspaces=false,                
  showstringspaces=false,
  showtabs=false,                  
  tabsize=2
}

\begin{document}

\title{ZS4C: Zero-Shot Synthesis of Compilable Code for Incomplete Code Snippets using LLMs} 


\author{Azmain Kabir}
\orcid{0009-0002-6077-7631}
\affiliation{%
  \institution{University of Manitoba}
  \city{Winnipeg}
  \country{Canada}}
\email{kabira1@myumanitoba.ca}

\author{Shaowei Wang}
\affiliation{%
  \institution{University of Manitoba}
  \city{Winnipeg}
  \country{Canada}}
\email{shaowei.wang@umanitoba.ca}

\author{Yuan Tian}
\affiliation{%
  \institution{Queen's University}
  \city{Kingston}
  \country{Canada}}
\email{y.tian@queensu.ca}

\author{Tse-Hsun Chen}
\affiliation{%
  \institution{Concordia University}
  \city{Montreal}
  \country{Canada}}
\email{peterc@encs.concordia.ca}

\author{Muhammad Asaduzzaman}
\affiliation{%
  \institution{University of Windsor}
  \city{Windsor}
  \country{Canada}}
\email{masaduzz@uwindsor.ca}

\author{Wenbin Zhang}
\affiliation{%
  \institution{Florida International University}
  \country{USA}}
\email{wenbinzhang2008@gmail.com}

\renewcommand{\shortauthors}{Kabir et al.}

\begin{abstract}

Technical Q\&A sites are valuable for software developers seeking knowledge, but the code snippets they provide are often uncompilable and incomplete due to unresolved types and missing libraries. This poses a challenge for users who wish to reuse or analyze these snippets. Existing methods either do not focus on creating compilable code or have low success rates. To address this, we propose ZS4C, a lightweight approach for zero-shot synthesis of compilable code from incomplete snippets using Large Language Models (LLMs). ZS4C operates in two stages: first, it uses an LLM, like GPT-3.5, to identify missing import statements in a snippet; second, it collaborates with a validator (e.g., compiler) to fix compilation errors caused by incorrect imports and syntax issues. We evaluated ZS4C on the StatType-SO benchmark and a new dataset, Python-SO, which includes 539 Python snippets from Stack Overflow across the 20 most popular Python libraries. ZS4C significantly outperforms existing methods, improving the compilation rate from 63\% to 95.1\% compared to the state-of-the-art SnR, marking a 50.1\% improvement. On average, ZS4C can infer more accurate import statements (with an F1 score of 0.98) than SnR, with an improvement of 8.5\% in the F1.

\end{abstract}

%

\keywords{Incomplete code snippets, Large Language Model, Program synthesis, API inference, Prompt engineering, ChatGPT}


\maketitle

\section{Introduction}

Technical question and answering sites (Q\&A) such as Stack Overflow have become an important source for software developers to seek knowledge, including code snippets that demonstrate API usage. A recent study revealed that 15 million posts on Stack Overflow have source code snippets, forming a vast codebase for developers to reuse~\cite{wu2019developers}. Recent surveys reported that developers search for code snippets in Q\&A sites frequently and extensively~\cite{sadowski2015developers,wu2019developers,xia2017developers}.

However, the majority of the code snippets on Q\&A sites are uncompilable, which raises the obstacle for users to reuse or analyze these code snippets~\cite{terragni2016csnippex,rubei2020postfinder}. Those code snippets are usually uncompilable and semantically incomplete for compilation due to unresolved types (e.g., what is the fully qualified name (FQN) for the type DateFormat?) and missing dependent libraries (e.g., which library to import for DateFormat?)~\cite{terragni2016csnippex,mishne2012typestate,dong2022snr}. Although missing such implementation details could be synthesized manually, it is tedious and often requires substantial familiarity with various libraries from developers. Furthermore, such manual synthesis effort is, however, not scalable for software engineering research based on Q\&A sites' code snippets~\cite{zhang2021study,zhang2019analyzing,gao2015fixing,mishne2012typestate}. Besides, Integrated Development Environments (IDEs) sometimes fail to provide accurate import statements for the code snippets~\cite{intellij_support}. This issue is also prevalent in LLM-generated code snippets, which can occasionally miss the appropriate import statements for successful execution~\cite{pan2024lost}. A technique to automatically repair code snippets by resolving the missing types and dependent libraries for compilation will significantly save developers time and facilitate the research based on such code snippets.

To address this issue, various approaches have been proposed~\cite{terragni2016csnippex,dong2022snr,saifullah2019learning,subramanian2014live,phan2018statistical}. For instance, CSNIPPEx synthesizes compilable code by inferring dependency libraries for code snippets based on a set of hand-crafted heuristics~\cite{terragni2016csnippex}. Baker~\cite{subramanian2014live} and \snr~\cite{dong2022snr} extract constraints from the abstract syntax tree (AST) of code snippets and use constraint solving to infer the FQN for missing types and their corresponding dependent libraries. StatType~\cite{phan2018statistical} and Coster~\cite{saifullah2019learning} build statistical models from existing compilable source code to predict FQN for reresolved types. However, Coster and StatTyper are not designed directly for synthesizing compilable code from code snippets, instead, it designed for FQN inference. For the approaches (e.g., Baker, CSNIPPEx, and \snr) that are designed for synthesizing compilable code, their reported compilation success rates are lingering below 80\%, it's evident that there's room for developing new tools to improve compilation rates. 
Thus, we aim to explore the possibility of designing a lightweight tool, capable of addressing compilation errors in incomplete code snippets, without relying on pre-defined rules and program analysis.

Large language models, such as ChatGPT, have demonstrated their effectiveness in various software engineering tasks~\cite{li2023nuances,surameery2023use,ma2023scope,xie2023chatunitest}, serving as a knowledge base for consultation. The two key steps for synthesizing compilable code from incomplete code snippets, inferring dependent libraries, and fixing potential compilation errors naturally align with the strength of LLM. 

In this study, we introduce \ourTool, an approach designed to perform zero-shot synthesis of compilable code from incomplete code snippets using LLM. \ourTool operates in two stages. In the first stage, \ourTool utilizes LLM (e.g., ChatGPT), to identify missing import statements for a given code snippet, leveraging our designed task-specific prompt template. We apply the widely adopted self-consistency method to improve the robustness of the inference. In the second stage, \ourTool employs iterative conversations between an LLM and a validator (e.g., compiler) to fix errors caused by incorrect import statements and syntax errors. Specifically, \ourTool uses a validator to provide feedback, e.g., compilation error logs, which serve as additional context for an LLM to correct the errors. Compared with prior approaches, \ourTool is lightweight and does not require any training data, any pre-defined rules or heuristics, or program analysis like constructing AST.

We thoroughly evaluated \ourTool on a widely used benchmark called StatType-SO consisting of 267 code snippets manually collected from Stack Overflow posts~\cite{phan2018statistical} and a new dataset called Python-SO that contains 539 Python code snippets collected from Stack Overflow across 20 popular Python libraries. \ourTool outperforms baselines on all studied datasets. We evaluated \ourTool with two models, GPT-3.5 and GPT-4.
\ourTool improves the compilation rate of the SOTA approach \snr, from 63\% to 95.1\%, with a 50.1\% improvement. On average, \ourTool can infer more accurate import statements than \snr, with an improvement of 8.5\% in the F1 score. Our ablation analysis shows that both components in \ourTool play a crucial role in the effectiveness of \ourTool. For instance, by employing the component conversation between a validator and GPT-3.5, \ourTool increases the compilation rate from 72\% to 87.3\% with a 21.3\% improvement. Our parameter analysis reveals that \ourTool is not sensitive to parameters. Lastly, we conducted a failure analysis and our results shed light on the limitations of using LLMs for synthesizing code, such as hallucination (e.g., fake FQN, and partially correct FQN) and unexpected code modification from LLMs. 

This paper makes the following contributions:
\begin{itemize}
    \item We propose \ourTool, a novel zero-shot LLM-based approach to automatically synthesize compilable code for code snippets, which employs iterative conversations LLM and validator (e.g., compiler) to boost the successful rate of synthesizing compilable code for code snippets.
    \item We conducted extensive experiments to demonstrate the effectiveness of \ourTool compared with the SOTA approach, and other baselines that mimic how developers use LLMs (e.g., ChatGPT) for resolving compilation issues for incomplete code snippets.
    \item We shed light on the limitation of using LLMs for the code synthesis task.
    \item We construct new data for Python libraries. We made our code and dataset publicly available~\cite{datarepo} to encourage future research on studying partial code synthesis and related approaches.
\end{itemize}

\section{Background and related work}
\subsection{Automated Synthesis of Compilable Code from Stack Overflow}\label{sec:Synthesis}

The task of automatically synthesizing compilable code snippets for incomplete code snippets on Q\&A is first introduced by Terragni et al. \cite{terragni2016csnippex}. They developed an Eclipse plug-in named CSNIPPEx for this task. The tool consists of three main components: C-Units Inference, Dependencies Resolver, and Code Completer. The C-Units Inference component is for resolving missing type declarations. The Dependencies Resolver component automatically identifies the necessary external libraries in their FQNs (fully qualified names, e.g., org.library1.sub.C)\footnote{Without fully qualified names, naming ambiguities can easily arise, as many simple class names across different libraries collide with one another.} based on a set of heuristics and generates import declarations. Finally, the Code Completer addresses issues like syntax errors, typos, and missing variable declarations by using Eclipse Quick Fix. A follow-up tool called APIzation was developed on top of CSNIPPEx to wrap up a code snippet and generate a reusable API with proper parameters, return statements, and missing external libraries~\cite{APIzation}. \ourTool draws inspiration from CSNIPPEx and consists of two components: one dedicated to identifying necessary library dependencies, and the other focused on iteratively improving the code by compiling and correcting errors. One main difference is that we utilize carefully designed prompts to leverage the power of LLM (ChatGPT) for the two sub-tasks and no pre-defined heuristics and program analysis (i.e., AST construction) are required.

Another work from Dong et al~\cite{dong2022snr}, in which they proposed a tool named SnR, that can automatically infer FQNs, recommend libraries, and create import statements for code snippets, eventually making code compilable.

SnR is a constraint-based approach. It first analyzes the incomplete code snippet and extracts constraints that capture the relationships between different elements in the code. Next, SnR utilizes a knowledge base built from a collection of real Java libraries which contains information about classes, methods, and dependencies. The extracted constraints are then solved using a Datalog solver called Soufflé~\cite{dong2022snr}. This solver uses the knowledge base to find solutions that satisfy the constraints. Once the constraints are solved, SnR transforms the solutions into import statements. These import statements are added to the code snippet to resolve missing dependencies. They have compared their tool with the state-of-the-art FQN inference approach named COSTER~\cite{saifullah2019learning}. Experiments on a benchmark of 267 code snippets from Stack Overflow show that SnR correctly infers 91.0\% of the import statements, which makes 73.8\% of the snippets compile, compared to 36.0\% of the import statements and 9.0\% of the snippets by COSTER~\cite{saifullah2019learning}. Given these results, we consider SnR as our baseline approach.

One of the key tasks in the automated synthesis of compilable code snippets is to identify the fully qualified name (FQN) for APIs (\textit{FQN inference})~\cite{huang2022prompt}. Specifically, identifying the types of API elements in incomplete code can help identify the dependent libraries. 

Existing studies addressed FQN inference in different ways: Baker extracts constraints from code snippets and uses a naive constraint solving algorithm to infer FQNs~\cite{subramanian2014live}; StatType formulates the task as a phrase-based machine translation task - translating the code snippet into a target sentence that contains the FQNs of the API elements~\cite{phan2018statistical} and builds a statistical model from existing compilable source code; COSTER~\cite{saifullah2019learning} is another statistical model-based approach, it calculates the likelihood of appearing context tokens and the FQN of each API element, and then ranks the FQNs based on three different scores: likelihood, context similarity, and name similarity scores; Huang et al.~\cite{huang2022prompt} treats FQN inference as clozy-style fill-in-blank language task. They utilized the pre-trained CodeBERT model~\cite{feng2020codebert} and fine-tuned it using prompt learning with FQN-annotated code examples, sampled from target libraries. The prompt learning stimulates the MLM to recognize FQN syntax and usage patterns. In our approach, we do not directly perform FQN inference, instead, we skip it and directly focus on import statement inference.

\subsection{Using LLMs for Software Engineering Tasks}

With the breakthrough of LLMs in the field of natural language processing, there's a burgeoning interest in investigating the potential of LLMs to enhance software engineering tasks. Researchers have pre-trained LLMs on publicly available code, resulting in the development of specialized code LLMs like CodeBERT~\cite{feng2020codebert}, CodeT5~\cite{wang2021codet5}, Codex~\cite{chen2021evaluating}, and CodeGen~\cite{nijkamp2022codegen}. In addition to these pre-trained code models, another research trajectory focuses on optimizing the application of LLMs to software engineering tasks such as code generation~\cite{liucodegen4libs}, program repair~\cite{wei2023copiloting,liu2024llm}, testing~\cite{schafer2023adaptive}, and beyond. Given the remarkable capability of LLMs to learn from zero or minimal examples, there's a growing exploration into how in-context learning and prompt engineering can elevate LLM-based applications in software engineering. For instance, Gao et al.~\cite{gao2023constructing} investigated the influence of the selection, sequence, and quantity of demonstration examples on in-context learning performance in code intelligence tasks, namely, code summarization, bug fixing, and program synthesis. They found that meticulously curated demonstrations can significantly outperform conventional demonstration construction techniques. Liu et al. introduced the LLM-CompDroid framework, which combines the strengths of LLMs and traditional tools for repairing configuration compatibility bugs in Android apps~\cite{liu2024llm}. Their experimental results demonstrate a significant enhancement in bug resolution performance and surpassing the state-of-the-art tool. Different from the aforementioned research, we focus on designing a new prompt engineering method to synthesize compilable code for incomplete code snippets on Q\&A sites by leveraging LLM as a 
knowledge base and collaborating with a compiler. We introduce ZS4C, a lightweight zero-shot approach that decomposes the task into two stages, and enables a collaborative conversation between the LLM and a compiler to resolve compilation errors.

\section{Methodology}\label{sec:method} 

\begin{figure}[ht]
\centering
\includegraphics[width=\linewidth]{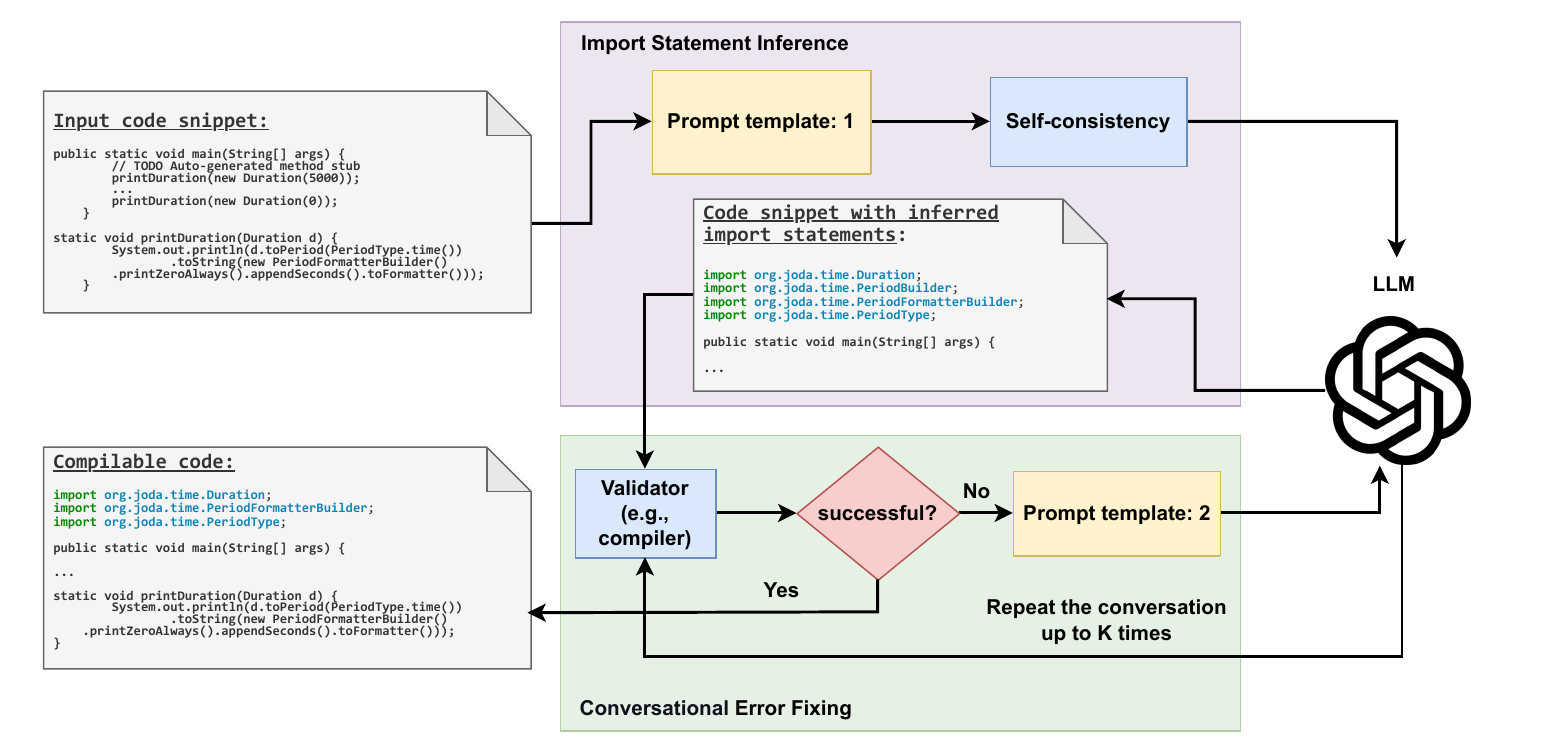}
\caption{Overview of our approach \ourTool. }
\label{fig:workflow}
\end{figure}

We present the overview of our approach \ourTool in Figure \ref{fig:workflow}. \ourTool is comprised of two components: \inferenceFullN and \compilationEnFullN. In \inferenceFullN, we leverage a large language model (LLM) to identify the missing dependent libraries (by identifying missing import statements) in the code snippet. In \compilationEnFullN, we further fix compilation errors caused by the wrong dependent libraries inferred by \inferenceFullN and other compilation errors (e.g., syntax error) to boost the compilation success rate of a code snippet, through the conversation between a validator such as a compiler which could provide error log, and the LLM. For simplicity, in the following section, we refer to \inferenceFullN as \inference and \compilationEnFullN as \compliationEn.

\subsection{\inferenceFullN}\label{sec:inference}
Previous studies report that the majority issue that prevents code snippets on Q\&A sites from compiling successfully is the absence of dependent libraries~\cite{terragni2016csnippex,dong2022snr}. Therefore, our first step is to infer the missing library import statements for a given code snippet. At a high level, \inference takes a code snippet $input\_code\_snippet$ as input and outputs the missing import statements for $input\_code\_snippet$. More specially, as shown in Figure~\ref{fig:workflow}, we first construct a prompt by using our designed simple yet effective prompt template. We then apply the self-consistency method~\cite{wang2022self} to obtain more reliable and coherent results from LLM. We elaborate on each step below.

\subsubsection{Prompt design}\label{sec:promptdesign}

Prompts are user-provided inputs, such as queries and instructions that guide LLMs and instruct their behavior for specific tasks. 
Previous studies have demonstrated that the quality of input prompts plays a crucial role in the performance of LLMs, significantly influencing the quality of the output~\cite{gao-etal-2021-making,white2023prompt,zhou2022large,min2022rethinking}. 
A prompt template is often used to generate prompts in a consistent and reproducible way. We follow the style of prior work~\cite{li2023nuances,xia2023conversational} in providing a short and concise prompt template tailored to the import statement inference task, as illustrated below.

\begin{tcolorbox}[colback=black!5!white,colframe=black!75!black,title=Prompt Template - 1 for \inferenceFullN]
{\footnotesize
  \underline{\textbf{\#Instruction}} \\
  \textit{Presentation\_Guide:} ``Reply with to-the-point answer, no elaboration.''\\
  
  \textit{Main\_Instruction (for Java):} ``Do not check for any import statements in the code. Only give correct imports by not using wildcard imports. Please note that you need to pay close attention and your response should be specific and accurate. Avoid repetition and must not generate wrong and nonexistent imports:'' \\

  \textit{Main\_Instruction (for Python):} ``Only give correct import statements for the attached code. Please note that you need to pay close attention and your response should be specific and accurate. Avoid repetition and must not generate wrong imports:'' \\
  
  \underline{\textbf{\#Query}} \\
 \{$input\_code\_snippet$\}
 }
\end{tcolorbox}

In \textit{Instruction}, we first use \textit{Presentation\_Guide} ``Reply with to-the-point answer, no elaboration'' to guide the model to provide a compact and to-the-point answer to our query. Without this directive, the model occasionally includes explanations in the output, which brings additional workload for extracting the missing import statements from the output. 

We design the instructions to guide and control the output of the model for our task step by step in \textit{Main\_Instruction}. We first ask the model to ignore any import statements in the given input code if present.\footnote{Note that, in line with previous studies~\cite{dong2022snr,phan2018statistical,saifullah2019learning}, the input code snippets utilized for evaluation in this study do not contain any import statements. However, code snippets on SO sometimes come with import statements.} This design directs the model’s attention exclusively to the body of the code rather than any existing import statements, which could be incorrect. Moreover, we observed that the model sometimes generates responses including ``no imports are found'', likely stemming from a misinterpretation of the task (e.g., to find import statements in the provided code snippet). To mitigate this issue, we explicitly instruct the model to avoid seeking import statements from the input code snippet. Another issue is wildcard import, which refers to the mechanism by which one can import all classes and interfaces from a particular package without specifying their fully qualified names by using
the asterisk (*) character. For instance, in Java, one can use ``import java.util.*;'' to import all classes from the ``java.util'' package, including the ArrayList, HashMap, and Scanner classes, among others, rather than individually importing classes, e.g.,  ``import java.util.ArrayList''. Wildcard may cause several issues: 1) wildcard imports may cause naming conflicts if multiple classes with the same name exist in different packages; 2) we want to steer the model towards identifying the exact classes required rather than resorting to the short-cut wildcard imports; 3) wildcard imports are generally discouraged in practice because it can compromise the code readability and introduce unintended side-effects that can be very difficult to
debug~\cite{van2021prevalence}. Therefore, we instruct the model to provide only correct imports and to avoid using wildcard imports. Following the recommended best practice for prompt design mentioned in ~\cite{white2023prompt}, we add a sentence ``Please note that you need to pay close attention and your response should be specific and accurate.'' to explicitly emphasize the importance of specificity and accuracy for the LLM to prioritize. Without this sentence, LLM sometimes outputs explanations with the inferred import statements and brings us challenges in extracting the import statement from the output. We also observed that in some cases, the model generates duplicate import statements and incorrect import statements that reference non-existent libraries (e.g., hallucination). To mitigate this issue, we added the sentence ``Avoid repetition and must not generate wrong and nonexistent imports'' to the prompt template.

Finally, we instantiate the \textit{\{$input\_code\_snippet$\}} with the actual code snippet and send the filled prompt to LLM for inference.

\subsubsection{Self-consistency}\label{sec:selfconsistence}
LLMs are nondeterministic by nature and would respond differently to the same input prompts~\cite{wang2022self,chatgptAPI}. Many LLMs, including ChatGPT, provide a \textit{temperature} hyper-parameter, which ranges from 0 to 1, to control the creativity or randomness of the generated response. A higher temperature (e.g., 0.7) results in more diverse and creative output while a lower temperature (e.g., 0.2) makes the output mode deterministic and focused. However, we observed that, even with a low temperature of 0, the import statements generated by LLM could vary time by time with the same prompt. For instance, using the same prompt for a Java class we got three different outputs:
1) ``The code does not require any additional import statements.'' 2) ``The code requires the following imports: java.text.NumberFormat; java.util.Locale''. 3) ``import java.text.NumberFormat;\\import java.util.Locale;''.

To obtain reliable and coherent results from the model, we employ the self-consistency method proposed by Wang et al.~\cite{wang2022self}. As indicated by its name, self-consistency simply asks a model to run the same prompt multiple times and determines the final answer based on the principle of frequency majority. More specifically, we run the same prompt for $K$ times and the model generates $K$ lists of import statements for our query. We count the frequency of each list and take the list with the highest frequency as the final result. If at least two lists have the same frequency, we repeat the process (i.e., running the same prompt for another $K$ times) until we break the tie. For instance, if we query the model using the same prompt 5 times, and the model generates five lists, $L_1 = \{A, B, C\}$, $L_2 = \{B, A, C\}$, $L_3 = \{A, B, C, D\}$, $L_4 = \{A, B\}$, and $L_5 = \{C, B, A\}$. We take $\{A, B, C\}$ as the final result as it (the combination of three libraries - A, B, C) appears 3 times with the highest frequency. In this study, we empirically set $K$ to 10, which achieves the best performance (see results of RQ3). Note that we only apply self-consistency in the Import Statement Inference component.  

\subsection{\compilationEnFullN}\label{sec:compilationFixing}

Through \inference, we could identify potential missing import statements. However, the code may still not be compiled due to incorrect import statements, bad syntax, typos, and undeclared variables. Therefore we need further validation of our generated code and boost the compilation rate by fixing those errors. Drawing inspiration from recent advancements in the Multi-Agent Conversation and its applications across various tasks~\cite{wu2023autogen,zhang2023steam,xi2023rise}, where multiple LLM-based agents work collaboratively and outperform a single LLM-based agent, we design a \compliationEn component to further improve the results given by \inference. Within this component, a validator (e.g., a compiler for Java code) is integrated into the workflow with the LLM (via conversations) to address compilation errors. This validator not only validates the generated code but also produces an error log that aids in resolving compilation issues. Specifically, we leverage the feedback, i.e., the error log, from the validator as context, prompting the LLM to fix the errors for the given code snippet.

\begin{algorithm}[h]
\footnotesize
    \SetKwFunction{Compile}{Compile}
    \SetKwFunction{queryLLM}{queryLLM}
    \SetKwFunction{constructPrompt}{constructPrompt}
    \SetKwInOut{KwIn}{Input}
    \SetKwInOut{KwOut}{Output}

    \KwIn{Code snippet with inferred import statements, \textbf{$CS$}}
    \KwIn{Maximum number of interaction between validator and LLM, \textbf{$M$}}
    \KwIn{Large language model, \textbf{$LLM$}}
    \KwOut{Most recent fixed Code, \textbf{$FC$}}
    $errorLog$ = "" \\
    $history$ = "" \\
    \tcp{Conversation between validator and LLM up to $K$ times}
    \For{$i \leftarrow 0$ \KwTo $M-1$}{
        $ValidationResults$ = Validator.validate($CS$)\\
        \tcp{$CS$ failed to pass the validator}
        \eIf{$ValidationResults$ is failed}{
            $errorLog$ = getErrorLog($ValidationResults$) \\
            $Prompt$ = constructPrompt($errorLog$, $history$) \\
            $CS$, $history$ = queryLLM($LLM$, $Prompt$) \\
          
         }{
           
            break
         }
    }
    $FC = CS$ \\
    \KwRet{$FC$} 
    \caption{Algorithm for \compilationEnFullN}\label{alg:compilation}
\end{algorithm}

We present the logic of \compliationEn in Algorithm~\ref{alg:compilation}. The algorithm takes the code snippet with inferred import statements generated by \inference, denoted as $CS$, the maximum number of interactions between validator and LLM, represented by $M$, and a large language model, $LLM$, as input, and outputs a most recent fixed code, denoted as $FC$. First, we examine if $CS$ can pass the validator successfully (line 5). If the validation is successful, we return $CS$ directly. If not, we construct a prompt incorporating the error log and $CS$ (lines 8 - 9). Subsequently, We query the $LLM$ with the constructed prompt, seeking corrections for $CS$ (line 10). This conversation (lines 4 - 14) is repeated up to $M$ times until a successful compilation.

 \begin{tcolorbox}[colback=black!5!white,colframe=black!75!black,title=Prompt Template - 2 for \compilationEnFullN]
{\footnotesize
\textit{Presentation\_Guide}: ``Reply with to-the-point answer, no elaboration.''
\\

\textit{Main\_Instruction (for Java)}: ``Now fix the error by focusing on fixing the import statements by not using wildcard imports and must not modify code body which means do not change anything inside the class. So, it can be successfully compiled and reply with full code.''
\\

\textit{Main\_Instruction (for Python)}: ``Now fix the error by focusing on fixing the import statements. So, it can be run successfully and reply with full code.''
}

\tcblower
{\footnotesize
\underline{\textbf{\# Prompt template for the 1st attempt:}}
\\
\{Presentation\_Guide\}
\\
``See the code below:'' \\
$\{input\_code\_snippet\_with\_inferred\_import\_statements\}$ \\
``For the above code I got the below error log:''\\
$\{error\_log\}$ \\
\{Main\_Instruction\}
\\\\
\underline{\textbf{\# Prompt starting from the 2nd attempt:}}
\\
\underline{$Context$:}\\
$\{chat\_history\}$
\\
$\{last\_code\_from\_ChatGPT\}$\\
``You gave the above imports fix in your attempt $\{attempt\_number\}$. But compiler gave this error:''\\
$\{last\_error\_message\}$\\
\{Main\_Instruction\}
}

\end{tcolorbox}

In each iteration, we update the $CS$ as suggested by $LLM$ and provide the chat history as contextual information. If, after attempting $M$ times, $CS$ remains unsuccessful, the latest revised code is returned. Our design mimics how developers use an LLM (e.g., ChatGPT) to fix compilation errors for a piece of code over conversation with LLM (e.g., ChatGPT).

We present our designed prompt template-2. We designed two templates: one for the first attempt and the other for subsequent attempt(s). In the first attempt, we attach the uncompilable code with an error log and ask LLM to fix the error and return a fixed version. Similar to the Prompt Template - 1 used in \inference (see Section~\ref{sec:inference}), we ask LLM to avoid wildcard imports. We observed cases where the LLM circumvents errors by simply removing code segments that cause the compilation error. To mitigate this issue, we instruct the LLM to not modify the code body. For the following attempt, we attach the history from the previous attempts and ask LLM to fix based on history. Since the size of LLM's context window is limited, we only keep the most recent chat history to offer the context that fits the window size.

Note that during the validation, if a compiler is used as a validator for programming languages such as Java and C++ which needs compilation first, the compiler needs to search for the external libraries to import as external dependencies. To do so, we dynamically import the external libraries based on the import statements obtained from \inference or previous round of \compliationEn. More specially, we construct a knowledge base to store the mappings between fully qualified names (FQN) of classes/functions and their corresponding external libraries (e.g., org.joda.time.Duration $\rightarrow$ joda-time-2.9.9.jar) in an offline manner. See more details of knowledge base construction in Section~\ref{sec:dataset}. When a piece of code with inferred import statements is input into \compliationEn, we search the external libraries for each FQN from the import statements and load the corresponding libraries from the constructed knowledge base. It's worth noting that a single FQN possibly could correspond to multiple external libraries (e.g., different versions of the same libraries provide the same class)~\cite{dong2022snr,subramanian2014live}. To mitigate such a case, we selected the one that satisfies pre-defined constraints following the previous study~\cite{subramanian2014live}. In our study, we consider the invoked functions and attributes as the constraints. If there exist multiple libraries available after resolving constraints, we select the one that was released most recently. We assume it is realistic for a user to use the most recent libraries. Note that our approach does not guarantee to find all external libraries due to various reasons (e.g., incorrect FQN in import statements), instead our goal is to leverage the message from the compiler to help LLM infer the correct import statements iteratively. For instance, the example in Figure~\ref{fig:rq2_before_after} (left), for the import statement ``import org.joda.time.PeriodFormatterBuilder;'', no external library was found since the import org.joda.time.PeriodFormatterBuilder (missing format) is incorrect. We still compile the code and the error message generated by the compiler helps LLM fix the error (see more detailed results in Section~\ref{sec:rq2}).

For Java code snippets, we select the standard compiler Conversational Error Fixing component. For Python code snippets, we use Pyflakes~\cite{pyflakes}, which is a static analysis tool for Python that checks for errors (syntax errors, undefined variables, and other common issues) in Python source files. For this study, we utilized Pyflakes version 3.2.0 to analyze our codebase before executing it with a Python interpreter.

\section{Experimental Design}

In this section, we present research questions (RQs), dataset, evaluation metrics, our analysis approach for RQs, and implementation details.

\subsection{Research questions}
We evaluate \ourTool in different aspects to answer the following research questions.
\begin{itemize}
    \item\rqone
    \hfill
    \item \rqtwo
    \hfill
    \item \rqthree
    \item \rqfour
\end{itemize}

\subsection{Data preparation}\label{sec:data}

\subsubsection{Datasets}\label{sec:dataset}
To evaluate the effectiveness of \ourTool, we first use an established benchmark called \textit{StatType-SO}, which was used in the related studies~\cite{dong2022snr,phan2018statistical,saifullah2019learning}. The dataset consists of 267 code snippets of six popular Java libraries. The six libraries are of different sizes and cover various application domains. This dataset was manually collected from Stack Overflow posts and represents the scenario in which developers use a wide variety of real Java libraries in practice~\cite{phan2018statistical}. This benchmark consists of the code snippets and their dependent libraries. 

To complement and diversify StatType-SO, which only consists of Java libraries, we also constructed a Python dataset following previous studies~\cite{phan2018statistical,subramanian2014live}. We first selected the top 20 most popular Python libraries~\cite{top30pythonlib}. For each library $L$ of interest, we searched for all the posts from the past 10 years that use $L$ via Stack Overflow tags on StackExchange Data Explorer\footnote{https://data.stackexchange.com/} and downloaded them. To ensure the quality of the posts, we only included those with a score greater than 5. We considered code snippets that have more than ten lines of code in both questions and answers. We ended up with 539 code snippets across the 20 studied libraries.

To construct the benchmark (i.e., the correct import statements for each Python code snippet), the first author manually collected the required modules by examining the posts on Stack Overflow that contain the corresponding code snippet and searching on Google. Note that some code snippets initially came with partially imported modules; we removed those modules when constructing the dataset to avoid data leakage.

Table~\ref{tab:dataset} summarizes the basic statistics of StatType-SO and Python-SO. 

\subsubsection{Knowledge base construction}
In Section~\ref{sec:compilationFixing}, we use a compiler to validate the code snippet and provide feedback (i.e., error log) to the LLM for error fixing for Java code snippets. To make the compiler ready for compilation, we need to import proper external libraries for dependencies. For this, we need to construct the knowledge base that stores the mapping between fully qualified names (FQN) of classes/functions and their corresponding external libraries. To do so, we downloaded all versions of the most popular 500 libraries from maven repositories\footnote{https://mvnrepository.com/popular}. We then performed the following steps to extract the FQN and the last modification date for each external library. 

\begin{itemize}
    \item\textbf{FQN Extraction:} We first decompress the jar file for each external library. Next, we use the \textit{javap} tool\footnote{https://docs.oracle.com/javase/8/docs/technotes/tools/windows/javap.html}, which is a Java class file disassembler, to analyze the binary class files extracted from the JAR. The output from javap is parsed to identify all non-abstract, non-interface class definitions, and then extract their fully qualified names.

    \item\textbf{Lastest Modification Date:} Note that we also need the modified date of each library to select the most recent library if multiple libraries meet constraints. To retrieve the last modification date of the JAR file, we examine the metadata associated with the file within the JAR. This date provides a timestamp indicating the most recent update or release of the JAR file. Note that for some jars we were not able to extract the valid modification date. Therefore, we added them manually from the Maven repository.
\end{itemize}

We then create the inverse index for each pair of FQN and its corresponding library (e.g., org.joda.time.Duration $\rightarrow$ joda-time-2.9.9.jar).

For Python code snippets, we do not use a compiler, since it is an interpreted language. Therefore, we did not construct a separate knowledge base. We use Pyflakes~\cite{pyflakes} to perform the validation and provide feedback in Section~\ref{sec:compilationFixing}. After that, we execute Python code by invoking the Python interpreter as a subprocess within our scripts. To ensure a consistent and isolated environment for running the Python code, we use a virtual environment. We identified and listed the necessary libraries for Python-SO, ensuring their installation in the latest versions within this virtual environment. This strategy enabled us to manage dependencies efficiently and avoid conflicts with other projects.

\begin{table}
    \centering
    \caption{Basic statistics of StatType-SO and Python-SO, including the number of code snippets, the average LOC, and unique import statement in each library. Note that \#Unique Import Statements refers to the total number of unique import statements that need to be predicted correctly for each library.}
    \label{tab:dataset}
    \footnotesize
    \begin{tabular}{|l|c|>{\centering\arraybackslash}m{0.7in}|>{\centering\arraybackslash}m{0.7in}|>{\centering\arraybackslash}m{1.0in}|>{\centering\arraybackslash}m{0.8in}|}
    \hline
       \multicolumn{6}{|c|}{\textbf{StatType-SO}}\\
       \hline
       \multicolumn{1}{|c|}{\textbf{Library}}  &  \textbf{\#Code Snippet} & \textbf{Mean \#LOC} & \textbf{Max \#LOC} & \textbf{Standard Deviation of \#LOC} & \textbf{\#Unique Import Statements} \\
       \hline
        Android & 50 & 35 & 192 & 34 & 129 \\
        \hline
       JDK  & 23 & 82 & 300 & 64 & 103 \\
       \hline
        Hibernate & 50 & 42 & 136 & 29 & 123\\
        \hline
        JodaTime   & 50 & 17 & 49 & 7 & 48 \\
        \hline
        GWT   & 50 & 31 & 111 & 16 & 108 \\
        \hline
        XStream  & 44 & 35 & 100 & 18 & 80 \\
        \hline
       \multicolumn{6}{|c|}{\textbf{Python-SO}}\\
       \hline
       \multicolumn{1}{|c|}{\textbf{Library}}  &  \textbf{\#Code Snippet} & \textbf{Mean \#LOC} & \textbf{Max \#LOC} & \textbf{Standard Deviation of \#LOC} & \textbf{\#Unique Import Statements} \\
       \hline
        BeautifulSoup & 35 & 13 & 29 & 6 & 31 \\
        \hline
        Flask & 23 & 14 & 29 & 6 & 32 \\
        \hline
        Gensim & 11 & 15 & 28 & 5 & 16 \\
        \hline
        Keras & 17 & 24 & 116 & 24 & 38 \\
        \hline
        Lightgbm & 7 & 18 & 36 & 11 & 13\\ 
        \hline
        Matplotlib & 49 & 13 & 36 & 7 & 18\\
        \hline
        NLTK & 18 & 14 & 46 & 9 & 26 \\
        \hline
        Numpy & 17 & 22 & 109 & 24 & 21 \\
        \hline
        OpenCV & 50 & 19 & 109 & 16 & 14 \\
        \hline
        Pandas & 8 & 18 & 36 & 10 & 9 \\
        \hline
        Plotly & 53 & 20 & 76 & 11 & 27 \\
        \hline
        Pytorch & 26 & 16 & 30 & 7 & 24 \\
        \hline
        Scikit-learn & 17 & 15 & 29 & 6 & 24 \\
        \hline
        Scipy & 37 & 16 & 65 & 10 & 44  \\
        \hline
        Scrapy & 49 & 19 & 113 & 18 & 74 \\
        \hline
        Seaborn & 48 & 15 & 53 & 10 & 17\\
        \hline
        Spacy & 26 & 14 & 26 & 4 & 26 \\
        \hline
        Tensorflow & 16 & 36 & 132 & 36 & 23\\
        \hline
        Theano & 11 & 11 & 18 & 3 & 20 \\
        \hline
        XGBoost & 21 & 21 & 75 & 14 & 35\\

        \hline
    \end{tabular}
   
\end{table}

\subsection{Evaluation metrics}\label{evaluation}

To evaluate the effectiveness of \ourTool as compared with alternatives, we consider the following two criteria.

\noindent \textbf{Criterion 1: Compilation Rate on the Synthesized Code.} To measure the effectiveness of \ourTool in synthesizing compilable code from incomplete code snippets, we use \textit{compilation rate} (CR), i.e., the ratio of code snippets that successfully compile post-repair to the total number of snippets evaluated, following previous studies~\cite{dong2022snr,terragni2016csnippex}. Note that Python code does not need to be compiled, therefore, we use a standard Python Interpreter to execute the code, and if it passes, it typically refers to the successful execution. For simplicity, throughout the paper, we consider it compilable and refer to it as the compilation rate (CR).

\noindent \textbf{Criterion 2: Accuracy of the Synthesized Import Statements.} A trivial fix, such as removing dependent libraries and related code can also resolve compilation errors. However this would risk significantly altering the code's semantics. Thus, in addition to the compilation rate, we assess the accuracy of the synthesized import statement list by comparing them to the ground truth list. To qualify the alignment between the two import statement lists, we define metrics \textit{True Positive (TP)}, \textit{False Positive (FP)}, \textit{False Negative (FN)} as:
    \begin{itemize}
    \item \textbf{True Positive (TP):} Denotes the count of import statements in the ground truth that are correctly predicted in the synthesized list.     
    \item \textbf{False Positive (FP):} Denotes the count of import statements in the synthesized list that are not present in the ground truth. 
    \item \textbf{False Negative (FN):} Denotes the count of import statements missing in the synthesized list but present in the ground truth.
\end{itemize}
Next, for each testing library, we aggregate the above values by summing the counts from each testing code snippet belonging to the library and calculate common classification metrics, i.e., Precision, Recall, and F1 as follows:
        $\text{Precision} = \frac{\text{TP}}{\text{TP} + \text{FP}}$, $
        \text{Recall} = \frac{\text{TP}}{\text{TP} + \text{FN}}$, $
        \text{F1} = 2 \times \frac{\text{Precision} \times \text{Recall}}{\text{Precision} + \text{Recall}}$. 

Following the baseline SnR~\cite{dong2022snr}, we also measure a list-wise matching degree for each library. Specifically, for each code snippet, the synthesized result is classified into one of the following categories:
\begin{itemize}
    \item \textbf{Same:} The approach synthesizes the exact same expected set of import statements.
    \item \textbf{Different:} The approach synthesizes one or more alternatives for some expected import statements.
    \item \textbf{Missing:} The approach synthesizes a subset of the expected set of import statements.
    \item \textbf{Extra:} The approach synthesizes a superset of the expected set of import statements.
    \item \textbf{None:} The approach synthesizes no import statements.
\end{itemize}
Finally, we summarize the distribution of the above categories for each testing library.

\subsection{Base large language model}
In this study, we choose GPT-3.5 and GPT-4 as our base LLMs~\cite{chatgpt}. We select them as they are recognized to be the most powerful LLMs and have gained wide attention in recent months by demonstrating the outstanding capability of solving tasks in various areas~\cite{frieder2023mathematical}, including software engineering~\cite{li2023nuances,surameery2023use,ma2023scope,xie2023chatunitest,xia2023conversational}. We accessed the ``gpt-3.5-turbo-0125'' for GPT-3.5 and ``gpt-4o'' model for GPT-4, via the OpenAI API~\cite{chatgptAPI, chatgpt4API}. We denote the \ourTool with GPT-3.5 and GPT-4 as \ourToolGPTThree and \ourToolGPTFour, respectively.

\subsection{Implementation details}

Experiments were conducted on a Linux server equipped with one Nvidia RTX 3090 GPU, a 24-Core CPU, and 64 GB of RAM. The implemented code has been made available on GitHub. We use the default setting (i.e., Maximum length = 256, Top P = 1, Frequency penalty = 0, and Presence penalty = 0) for ChatGPT except for temperature. For self-consistency, we set a high temperature of 1 following the recommendation by previous study~\cite{wang2022self}. For other components that involve ChatGPT, we set a medium temperature of 0.5.

\subsection{Approach of RQs}
\subsubsection{Approach of RQ1}
We compare \ourTool with the following two baselines:
\begin{itemize}
    \item \textbf{\snr:} The state-of-the-art technique for synthesizing incomplete code snippets by inferring missing import statements and making them compilable~\cite{dong2022snr}. \snr has been demonstrated to archive better performance than other approaches (e.g., COSTER~\cite{saifullah2019learning}, Baker~\cite{subramanian2014live}, and StatType~\cite{phan2018statistical}) on StatType-SO. Therefore, in this study, we only compare with \snr. We leveraged the replication package of \snr as provided in their paper.\footnote{We found the disparity between the number reported in the paper and the number derived from their replication package. We present the numbers derived from their replication package in this paper.} We do not compare with CSNIPPEx because this Eclipse plugin is no longer accessible.
    \item \textbf{\basicprompt:} We also consider a baseline that employs a straightforward prompt to instruct an LLM to make the given code snippet compilable. Specifically, we replace the Main\_Instruction as ``Make the code below compilable: \{$input\_code\_snippet$\}'' in Prompt Template-1 (ref. Section~\ref{sec:inference}) and while retaining the remainder of the prompt template unchanged. We denote the \basicprompt with GPT-3.5 and GPT-4 as \basicpromptGPTThree and \basicpromptGPTFour, respectively.
\end{itemize}

We evaluate the effectiveness of \ourTool with GPT-3.5 and GPT4 against the two baselines on the two criteria outlined in Section~\ref{evaluation} on StatType-SO. Since SnR is designed for Java and is not appliable to Python code, we focus on comparing \ourTool with \basicprompt for Python-SO dataset. To ensure the reliability of our results and mitigate potential biases arising from the inherent randomness of the LLM, we run both \ourTool and \basicprompt five times and take the median value as the final result.

\subsubsection{Approach of RQ2}
In this RQ, we carry out an ablation study to understand the contribution of each component (i.e., \compliationEn and \inference) in \ourTool. Note that for \compliationEn, the input code snippet will be the original code snippet without any inferred import statements. We are also interested in understanding the contribution of the self-consistency method using in \inference. Therefore, we remove self-consistency from \inference (refer as to \inferenceWOSC) and compare it with \inference. We then evaluate the effectiveness of the above three approaches following the same methodology in RQ1, and compare them with \ourTool. Note that using the API of GPT-4 is expensive, and we observe similar performance between GPT-4 and GPT-3.5. Therefore, we carry out the ablation study with the more economical model GPT-3.5.

\subsubsection{Approach of RQ3}
\ourTool has two parameters: $K$ - the number of sampling in self-consistency, and $M$ - the maximum rounds of conversation between the compiler and the LLM in \compilationEnFullN. In this RQ, we investigate the impact of each parameter on the effectiveness of \ourTool, where we set $K$ to 10, 20, and 30 and set $M$ to 5, 10, and 15, respectively. Similar to RQ2, we carry out the ablation study with GPT-3.5.

\subsubsection{Approach of RQ4}

Few-shot in-context learning is a paradigm where a limited set of examples are provided in the prompt and allows an LLM to learn from the examples in a specific context. This method has demonstrated its effectiveness in enhancing LLM performance, especially when compared to scenarios where no examples (zero-shot) are provided for various downstream tasks~\cite{nashid2023retrieval,brown2020language,gao-etal-2021-making,min2022rethinking}. Therefore, in this RQ, we aim to investigate if enabling this option would benefit \ourTool.  

Selecting good demonstrations is essential for few-shot in-context learning. Prior studies show that selecting demonstration examples with higher similarity (to the input) can improve in-context learning performance~\cite{nashid2023retrieval,zhou2022docprompting}. Thus, in this RQ, we incorporate code snippets that are most similar to each input code snippet, along with their corresponding import statements, as demonstrations within the prompt. Specifically, we employ the commonly used CodeBERT~\cite{feng2020codebert} model that was trained for code-related tasks to learn a vector representation for each code snippet in StatType-SO. Based on the code representations output by CodeBERT, we select the code snippets presenting the highest cosine similarities with the input code snippet. When testing the performance of ZS4C with few-shot in-context learning, we selected different configurations based on the capabilities of the language models and their token window size limitations. Table~\ref{tab:prompt_token_summary} presents the five-number summary of the prompt token size for different settings. As we can see the median size for three-shot prompts is 7,100, which indicates that the majority of the prompts exceed the window size of GPT-3.5. Therefore, for GPT-3.5, we utilized a one-shot learning approach, while for GPT-4, we employed both one-shot and three-shot learning approaches. This decision was made due to the token window size constraints (i.e., 4k input tokens for GPT-3.5 and 128k input tokens for GPT-4). We put the retrieved demonstrated example after the Main\_Instruction in Prompt Template - 1 as shown in Section~\ref{sec:promptdesign}.

\begin{table}[H]
\caption{Summary of prompt size for different settings in terms of input token size distributions, including the five-number summary statistics (Min: Minimum, Q1: 25th percentile, Median: 50th percentile, Q3: 75th percentile, Max: Maximum).}
    \centering
    \begin{tabular}{|p{0.7in}|p{0.25in}|p{0.25in}|p{0.4in}|p{0.32in}|p{0.38in}|p{0.25in}|p{0.32in}|p{0.4in}|p{0.32in}|p{0.4in}|}
    \hline
         \multicolumn{1}{|c|}{} & \multicolumn{5}{c|}{\textbf{StatType-SO}} & \multicolumn{5}{c|}{\textbf{Python-SO}} \\
        \cline{2-11}
        & \textbf{Min} & \textbf{Q1} & \textbf{Median} & \textbf{Q3} & \textbf{Max} & \textbf{Min} & \textbf{Q1} & \textbf{Median} & \textbf{Q3} & \textbf{Max} \\
        \hline
        \textbf{Three-shot} & 4,920 & 5,654 & 7,100 & 10,380 & 100,650 & 2,370 & 4,200 & 5,790 & 8,430 & 42,060 \\
        \hline
        \textbf{One-shot} & 1,200 & 1,934 & 3,320 & 4,840 & 65,020 & 1,580 & 2,800 & 3,860 & 5,620 & 28,040 \\
        \hline
        \textbf{Zero-shot} & 620 & 1,082 & 1658 & 2,440 & 32,530 & 790 & 1,400 & 1,930 & 2,810 & 14,020 \\
        \hline
    \end{tabular}
    \label{tab:prompt_token_summary}
\end{table}

\section{Results}
\label{sec:results}

\subsection{\rqone}\label{sec:rq1}

\begin{table}
\caption{The comparison of \emph{SnR}, \basicprompt, \ourTool with GPT-3.5 and \ourTool with GPT-4 for StatType-SO (Java) dataset in terms of F1, recall (Rec), precision (Pre), and compilation rate (CR). We present the average for F1, Rec, Pre, and the total compilation rate in the summary row. The cells that have the best performance are marked in bold.
}\label{tab:overallResults}
\scriptsize
\resizebox{\textwidth}{!}{%
\begin{tabular}
{|p{0.46in}|p{0.16in}|p{0.16in}|p{0.16in}|p{0.3in}|p{0.16in}|p{0.16in}|p{0.16in}|p{0.3in}|p{0.16in}|p{0.16in}|p{0.16in}|p{0.3in}|p{0.16in}|p{0.16in}|p{0.16in}|p{0.3in}|p{0.16in}|p{0.16in}|p{0.16in}|p{0.3in}|}
\hline
\multicolumn{1}{|l|}{\multirow{2}{*}{\textbf{}}} & \multicolumn{4}{c|}{\textbf{\snr}} & \multicolumn{4}{c|}{\textbf{\basicpromptGPTThree}} & \multicolumn{4}{c|}{\textbf{\basicpromptGPTFour}} & \multicolumn{4}{c|}{\textbf{\ourToolGPTThree}} & \multicolumn{4}{c|}{\textbf{\ourToolGPTFour}}
\\ \cline{2-21} 
\multicolumn{1}{|l|}{}                                  & \textbf{F1} & \textbf{Rec} & \textbf{Pre} & \textbf{CR}  & \textbf{F1} & \textbf{Rec} & \textbf{Pre} & \textbf{CR} & \textbf{F1} & \textbf{Rec} & \textbf{Pre} & \textbf{CR}  & \textbf{F1} & \textbf{Rec} & \textbf{Pre} & \textbf{CR} & \textbf{F1} & \textbf{Rec} & \textbf{Pre} & \textbf{CR}\\ 
\hline
\textbf{Android}   & 0.952 & 0.909 & 1 & 37 (74\%)         & 0.967 & 0.95   & 0.987 & 44 (88\%)             & 0.969 & 0.956 & 0.983 & 45 (90\%) & 0.990 & 0.990 & 0.990 & 48 (96\%) & 0.988 & 0.99 & 0.987 & 48 (96\%)        \\
\hline
\textbf{JDK}        & 0.887 & 0.796 & 1 & 12 (52.2\%)          & 0.482 & 0.317  & 1     & 15 (65.2\%)           & 0.713 & 0.557 & 0.989 & 18 (78.3\%) & 0.997 & 1.000 & 0.994 & 23 (100\%) & 0.994 & 1 & 0.988 & 23 (100\%)        \\
\hline
\textbf{Hibernate}  & 0.948 & 0.918 & 0.98 & 23 (46\%)         & 0.75  & 0.606  & 0.985 & 22 (44\%)            & 0.834 & 0.747 & 0.944 & 29 (58\%) & 0.902 & 0.834 & 0.982 & 32 (64\%) & 0.956 & 0.916 & 1 & 41 (82\%)          \\
\hline
\textbf{Joda-Time}  & 0.86 & 0.759 & 0.992 & 37 (74\%)         & 0.979 & 0.971  & 0.988 & 44 (88\%)            & 0.991 & 0.994 & 0.988 & 49 (98\%) & 0.985 & 0.994 & 0.977 & 50 (100\%) & 0.994 & 0.988 & 1 & 50 (100\%)          \\
\hline
\textbf{GWT}        & 0.835 & 0.721 & 0.991 & 30 (60\%)          & 0.951 & 0.927  & 0.977 & 32 (64\%)           & 0.906 & 0.889 & 0.924 & 42 (84\%) & 0.968 & 0.94 & 0.997 & 46 (92\%) & 0.987& 0.978 & 0.997 & 50 (100\%)          \\
\hline
\textbf{XStream}    & 0.934 & 0.877 & 1 & 29 (65.9\%)          & 0.863 & 0.78   & 0.967 & 22 (50\%)           & 0.927 & 0.89 & 0.967 & 38 (86.4\%) & 0.934 & 0.89 & 0.98 & 35 (79.6\%) & 0.959 & 0.925 & 0.995 & 42 (95.5\%)          \\
\hline
\textbf{Summary}    & 0.903 & 0.83 & 0.994 & 168 (62.9\%) & 0.832 & 0.759 & 0.984 & 179 (67\%)  & 0.890 & 0.839 & 0.966 & 221 (82.8\%) & 0.963 & 0.938 & 0.987 & 234 (87.6\%) & \textbf{0.98} & \textbf{0.966} & \textbf{0.995} & \textbf{254 (95.1\%)}\\
\hline
\end{tabular}%
}
\end{table}

\begin{table}[]\caption{The comparison of \basicprompt (with GPT-3.5 and GPT-4) and \ourTool (with GPT-3.5 and GPT-4) for Python dataset (Python-SO) in terms of F1, recall (Rec), precision (Pre), and compilation rate (CR). We present the average for F1, Rec, Pre, and the total compilation rate in the summary row. The cells that have the best performance are marked in bold. 
}\label{tab:overallResultsPython}
\scriptsize
\resizebox{\textwidth}{!}{%
\begin{tabular}{|p{0.8in}|p{0.16in}|p{0.16in}|p{0.16in}|p{0.3in}|p{0.16in}|p{0.16in}|p{0.16in}|p{0.3in}|p{0.16in}|p{0.16in}|p{0.16in}|p{0.3in}|p{0.16in}|p{0.16in}|p{0.16in}|p{0.3in}|}
\hline
\multicolumn{1}{|l|}{\multirow{2}{*}{\textbf{}}} & \multicolumn{4}{c|}{\textbf{\basicpromptGPTThree}}                                                                                                                                        & \multicolumn{4}{c|}{\textbf{\basicpromptGPTFour}}                                                                                                                                        & \multicolumn{4}{c|}{\textbf{\ourToolGPTThree}}                                                                                                             & \multicolumn{4}{c|}{\textbf{\ourToolGPTFour}}                                                                                                             \\ \cline{2-17} 
\multicolumn{1}{|l|}{}                                  & \textbf{F1} & \textbf{Rec} & \textbf{Pre} & \textbf{CR}  & \textbf{F1} & \textbf{Rec} & \textbf{Pre} & \textbf{CR}  & \textbf{F1} & \textbf{Rec} & \textbf{Pre} & \textbf{CR}  & \textbf{F1} & \textbf{Rec} & \textbf{Pre} & \textbf{CR}  \\ 
\hline
\textbf{BeautifulSoup}   & 0.954 & 0.913 & 1   & 29 (82.9\%)          & 0.947 & 0.9 & 1 & 31 (88.6\%)          & 0.968 & 0.938 & 1   & 31 (88.6\%)          & 0.974 & 0.95  & 1   & 33 (94.3\%)          \\
\hline
\textbf{Flask}           & 0.937 & 0.908 & 0.967 & 15 (65.2\%)         & 0.977 & 0.985 & 0.97 & 18 (78.3\%)               & 1     & 1     & 1   & 23 (100\%)          & 1     & 1     & 1   & 23 (100\%)          \\
\hline
\textbf{Gensim}          & 0.919 & 0.895 & 0.944 & 10 (90.9\%)         & 0.842 & 0.842 & 0.842 & 9 (81.8\%)               & 1     & 1     & 1   & 11 (100\%)          & 0.973 & 0.947 & 1   & 11 (100\%)          \\
\hline
\textbf{Keras}           & 0.947 & 0.953 & 0.942 & 15 (88.2\%)         & 0.948 & 0.965 & 0.932 & 13 (76.5\%)               & 0.988 & 0.988 & 0.988 & 14 (82.4\%)       & 0.994 & 0.988 & 1   & 15 (88.2\%)          \\
\hline
\textbf{LightGBM}        & 0.947 & 0.9   & 1    & 6 (85.7\%)          & 0.976 & 1 & 0.952 & 6 (85.7\%)               & 1     & 1     & 1   & 7 (100\%)           & 1     & 1     & 1   & 7 (100\%)           \\
\hline
\textbf{Matplotlib}      & 0.983 & 0.978 & 0.989 & 49 (100\%)         & 0.989 & 0.989 & 0.989 & 49 (100\%)               & 0.994 & 0.989 & 1   & 46 (93.9\%)          & 0.994 & 0.989 & 1   & 47 (95.9\%)          \\
\hline
\textbf{NLTK}            & 0.968 & 0.957 & 0.978 & 10 (55.6\%)         & 0.957 & 0.957 & 0.957 & 13 (72.2\%)               & 1     & 1     & 1   & 18 (100\%)          & 0.978 & 0.957 & 1   & 18 (100\%)          \\
\hline
\textbf{Numpy}           & 0.887 & 0.796 & 1    & 15 (88.2\%)         & 0.887 & 0.796 & 1 & 17 (100\%)               & 0.826 & 0.704 & 1   & 16 (94.1\%)          & 0.839 & 0.722 & 1   & 17 (100\%)          \\
\hline
\textbf{OpenCV}          & 0.984 & 0.989 & 0.978 & 25 (50\%)         & 0.989 & 1 & 0.979 & 31 (62\%)               & 1     & 1     & 1   & 48 (96\%)          & 1     & 1     & 1   & 49 (98\%)          \\
\hline
\textbf{Pandas}          & 0.973 & 0.947 & 1    & 7 (87.5\%)          & 0.947 & 0.947 & 0.947 & 8 (100\%)               & 0.973 & 0.947 & 1   & 7 (87.5\%)           & 0.973 & 0.947 & 1   & 7 (87.5\%)           \\
\hline
\textbf{Plotly}          & 0.92  & 0.886 & 0.956 & 24 (45.3\%)         & 0.946 & 0.919 & 0.974 & 27 (50.9\%)               & 0.94  & 0.887 & 1   & 40 (75.5\%)          & 0.937 & 0.896 & 0.982 & 42 (79.2\%)      \\
\hline
\textbf{Pytorch}         & 0.945 & 0.968 & 0.923 & 23 (88.5\%)         & 0.903 & 0.903 & 0.903 & 25 (96.2\%)               & 0.984 & 0.968 & 1   & 26 (100\%)          & 0.949 & 0.903 & 1   & 26 (100\%)          \\
\hline
\textbf{Scikit-learn}    & 1     & 1     & 1    & 17 (100\%)         & 1 & 1 & 1 & 17 (100\%)               & 1     & 1     & 1   & 17 (100\%)          & 0.989 & 0.979 & 1   & 17 (100\%)          \\
\hline
\textbf{Scipy}           & 0.963 & 0.948 & 0.979 & 24 (64.9\%)         & 0.958 & 0.938 & 0.978 & 26 (70.3\%)               & 0.974 & 0.949 & 1   & 37 (100\%)          & 0.979 & 0.959 & 1   & 37 (100\%)          \\
\hline
\textbf{Scrapy}          & 0.875 & 0.875 & 0.875 & 32 (65.3\%)         & 0.877 & 0.892 & 0.863 & 34 (69.4\%)               & 0.966 & 0.934 & 1   & 47 (95.9\%)          & 0.97  & 0.942 & 1   & 49 (100\%)          \\
\hline
\textbf{Seaborn}         & 0.974 & 0.974 & 0.974 & 45 (93.7\%)         & 0.965 & 0.974 & 0.957 & 47 (97.9\%)               & 0.987 & 0.974 & 1   & 48 (100\%)          & 0.987 & 0.974 & 1   & 48 (100\%)          \\
\hline
\textbf{Spacy}           & 0.926 & 0.875 & 0.982 & 18 (69.2\%)         & 0.935 & 0.906 & 0.967 & 25 (96.2\%)               & 0.984 & 0.969 & 1   & 16 (61.5\%)          & 0.984 & 0.969 & 1   & 23 (88.5\%)          \\
\hline
\textbf{Tensorflow}      & 0.961 & 0.961 & 0.961 & 7 (43.8\%)         & 0.949 & 0.922 & 0.979 & 10 (62.5\%)               & 0.99  & 0.98  & 1   & 13 (81.3\%)          & 1     & 1     & 1   & 15 (93.8\%)          \\
\hline
\textbf{Theano}          & 0.972 & 0.946 & 1    & 9 (81.8\%)          & 0.946 & 0.946 & 0.946 & 8 (72.7\%)               & 0.974 & 0.974 & 1   & 8 (72.7\%)          & 0.986 & 0.973 & 1   & 10 (90.9\%)          \\
\hline
\textbf{XGBboost}        & 0.971 & 0.944 & 1    & 17 (81\%)         & 0.965 & 0.958 & 0.971 & 18 (85.7\%)               & 0.986 & 0.972 & 1   & 18 (85.7\%)          & 0.986 & 0.972 & 1   & 20 (95.2\%)          \\
\hline
\textbf{Summary}         & 0.95 & 0.931 & 0.972 & 397 (73.7\%) & 0.945 & 0.937 & 0.955 & 432 (80.2\%) & \textbf{0.977} & \textbf{0.959} & \textbf{0.999} & 491 (91.1\%) & 0.975 & 0.953 & \textbf{0.999} & \textbf{514 (95.4\%)} \\
\hline
\end{tabular}%
}
\end{table}

\begin{table}
\footnotesize
\caption{The comparison of \snr, \basicprompt, \ourTool in terms of list-wise matching degree on StatType-SO and Python-SO datasets.}\label{tab:rq1_rq2_match}
\begin{tabular}{|l|c|c|c|c|c|}
\hline
\multicolumn{6}{|c|}{\textbf{StatType-SO}} \\
\hline
\multicolumn{1}{|c|}{\textbf{Approach}}                         & \textbf{Same} & \textbf{Different} & \textbf{Extra} & \textbf{Missing} & \textbf{None} \\
\hline

\textbf{\snr} & 143 (53.56\%) & 78 (29.21\%) & 4 (1.5\%) & 35 (13.11\%) & 7 (2.62\%) \\
\hline
\textbf{\basicpromptGPTThree} & 173 (64.79\%) & 55 (20.60\%) & 14 (5.24\%) & 19 (7.12\%) & 6 (2.25\%) \\
\hline
\textbf{\basicpromptGPTFour} & 226 (84.64\%) & 24 (8.99\%) & 10 (3.75\%) & 7 (2.62\%) & 0 (0\%) \\
\hline
\textbf{\ourToolGPTThree} & 202 (75.66\%) & 35 (13.11\%) & 16 (5.99\%) & 14 (5.24\%) & 0 (0\%) \\
\hline
\textbf{\ourToolGPTFour} & 226 (84.64\%) & 24 (8.99\%) & 10 (3.75\%) & 7 (2.62\%) & 0 (0\%) \\
\hline

\multicolumn{6}{|c|}{\textbf{Python-SO}} \\
\hline
\multicolumn{1}{|c|}{\textbf{Approach}}                         & \textbf{Same} & \textbf{Different} & \textbf{Extra} & \textbf{Missing} & \textbf{None} \\
\hline

\textbf{\basicpromptGPTThree} & 447 (82.93\%) & 43 (7.98\%) & 27 (5.01\%) & 19 (3.53\%) & 3 (0.56\%) \\
\hline
\textbf{\basicpromptGPTFour} & 447 (82.93\%) & 42 (7.79\%) & 39 (7.24\%) & 9 (1.67\%) & 2 (0.37\%) \\
\hline
\textbf{\ourToolGPTThree} & 491 (91.09\%) & 35 (6.49\%) & 4 (0.74\%) & 9 (1.67\%) & 0 (0\%) \\
\hline
\textbf{\ourToolGPTFour} & 499 (92.58\%) & 30 (5.57\%) & 3 (0.56\%) & 7 (1.30\%) & 0 (0\%) \\
\hline

\end{tabular}
\end{table}

We present the comparison results for StatType-SO in Tables~\ref{tab:overallResults} and for Python-SO in Table~\ref{tab:overallResultsPython}. The results show that both \ourToolGPTThree and \ourToolGPTFour outperform the baseline SnR by a large margin on StatType-SO. For instance, \ourToolGPTThree successfully synthesized 234 compilable code snippets out of the 267 provided incomplete code snippets, resulting in an impressive overall compilation rate of 87.6\%, outperforming two baselines. Compared with \snr, \ourToolGPTThree showed a 39.3\% improvement in terms of compilation rate.\footnote{SnR paper reports that they can compile 197 out of 267 (73.8\%) code snippets successfully after the repair. However, in their replication package, we found 168 code snippets were marked as comparable. Nevertheless, our approach outperforms SnR by improving the number of compilable code snippets to 234.} Furthermore, \ourToolGPTThree consistently outperformed the two approaches for all libraries. \ourToolGPTFour achieves a compilation performance improvement of 8.56\% over its predecessor GPT-3.5, increasing from 87.6\% to 95.1\%. Similarly, on Python-SO, \ourTool outperforms \basicprompt. For instance, when using GPT3.5 as the base model, \ourTool achieves a compilation rate of 91.1\%

When measuring the accuracy of the synthesized import statements, \ourTool always outperforms baselines in all evaluation metrics. For instance, on average, \ourToolGPTFour achieves 0.98, 0.966, and 0.995 in terms of F1, Precision, and Recall on StatType-SO, respectively, as detailed in Table~\ref{tab:overallResults}. Compared with \snr, \ourToolGPTFour achieves an improvement of 8.5\% and 16.4\% in terms of F1 and recall. \ourToolGPTFour and \snr have similar precision. In six of the target libraries, \ourToolGPTFour outperforms \snr in all six libraries. We perform failure cases analysis to understand why eventually some cases are not compilable after the fixing (see more details in Section~\ref{sec:failure}). 

Compared with \basicprompt, \ourTool consistently can predict import statements more accurately and compile more code snippets. For instance, on StatType-SO \ourToolGPTThree can compile 55 more code snippets, reflecting a 30.7\% enhancement in the compilation rate, compared with \basicpromptGPTThree. Furthermore, \ourToolGPTThree elevates the accuracy of import statement inference, with the F1 score rising from 0.832 to 0.961. These improvements indicate the significance of task-specific prompt engineering in optimizing the performance of LLMs. 
We observe a similar trend on Python-SO dataset. \ourTool consistently outperforms \basicprompt with the same base model. 

To complement accuracy metrics, we also compare the list-wise matching degree and present the result in Table~\ref{tab:rq1_rq2_match}. Both \ourToolGPTFour (226) and \ourToolGPTThree (202) can infer more exact match import statement lists (same) than \snr \footnote{SnR paper reports that they can achieve 183 same, 62 different, 11 extra, 4 missing, and 7 none matches. However, in their replication package, we found the number we reported in the table. Nevertheless \ourTool with GPT-3.5 can achieve more exact same matches.} and \basicprompt (193) on StatType-SO. We observe a similar trend on Python-SO, \ourTool always outperforms \basicprompt. 
Also \ourTool can always output things that are related to the correct import statements (none is 0). 

To understand why \ourTool outperforms \snr, we manually study the cases where \ourTool performs better than \snr. Since \ourToolGPTFour and \ourToolGPTThree achieve similar performance, here we study the cases from \ourToolGPTThree. We observe two reasons. First, \snr needs to repair an input code snippet to a syntactically valid compilation unit and then generate the Abstract Syntax Tree (AST) for the following procedures. However, the syntax repairing is not 100\% guaranteed. Therefore, in certain cases, \snr failed in the first repairing step. While \ourTool with GPT-3.5 does not rely on static analysis (e.g., parsing AST). LLM directly takes code snippets as sequence input and does the inference, which reduces the risk of failures due to errors incurring in static analysis. Second, \snr infers the appreciated import statements by solving constraints. However, if multiple libraries meet the constraints simultaneously, \snr does not have a strategy to handle this. While \ourTool leverages LLMs as the knowledge base to perform the inference, LLMs have sufficient knowledge to decide which libraries are most appreciated given the context.

In addition, we observe that \ourToolGPTFour consistently outperforms \ourToolGPTThree or achieves comparable effectiveness as \ourToolGPTThree on both studied datasets, typically in terms of compilation rate. \ourToolGPTFour improves the compilation rate of \ourToolGPTThree from 87.6\% to 95.1\%. This probably originates from two advantages of GPT-4. First, GPT-4 is more advanced than GPT-3.5 and has already proved more effective in various code-related tasks~\cite{liu2024your,poldrack2023ai}. Second, GPT-4 has a long window size and can fit much more context in the prompt (e.g., history generated during Conversational Error Fixing).

\rqbox{In general, \ourTool with GPT-4 outperforms GPT-3.5 and other baselines on both StatType-SO and Python-SO. For instance, on StatType-SO, \ourTool with GPT-4 improves the compilation rate of the SOTA approach, \snr, from 63\% to 95.1\%. On average, \ourTool's (with GPT-4) synthesized import statements are more accurate (with an F1 score of 0.98) than \snr, with an improvement of 8.5\% in the F1 score. Furthermore, \ourTool is more effective than \basicprompt.}

\subsection{\rqtwo}\label{sec:rq2}

\begin{table}[H]\caption{The comparison of \inferenceWOSC, \inference, \compliationEn in terms of F1, recall (Rec), precision (Pre), and compilation rate (CR) with GPT-3.5. We present the average for F1, Rec, and Pre and the total compilation rate in the summary row.
}
\scriptsize 
\begin{tabular}{|l|p{0.18
in}|p{0.22
in}|p{0.22
in}|p{0.3
in}|p{0.22
in}|p{0.22
in}|p{0.22
in}|p{0.3
in}|p{0.22
in}|p{0.22
in}|p{0.22
in}|p{0.3
in}|}
\hline
\multicolumn{1}{|l|}{\multirow{2}{*}{\textbf{}}} & \multicolumn{4}{c|}{\textbf{\inferenceWOSC}}                                                                                                                                               & \multicolumn{4}{c|}{\textbf{\inference}}                                                                                                                     & \multicolumn{4}{c|}{\textbf{\compliationEn}}                                                                                                                           \\ \cline{2-13} 
\multicolumn{1}{|l|}{}                                  & \textbf{F1} & \textbf{Rec} & \textbf{Pre} & \textbf{CR}  & \textbf{F1} & \textbf{Rec} & \textbf{Pre} & \textbf{CR} &  \textbf{F1} & \textbf{Rec} & \textbf{Pre} & \textbf{CR} \\ 
\hline
\textbf{StatType-SO}    & 0.923 & 0.893 & 0.960 & 193 (72.2\%)  & 0.951 & 0.924 & 0.984 & 205 (76.8\%) &  0.897 & 0.851 & 0.954 & 225 (84.2\%) \\
\hline
\textbf{Python-SO}    & 0.956  & 0.937 & 0.978 & 417 (77.4\%)  & 0.965 & 0.942 & 0.993 & 451 (83.7\%) & 0.942 & 0.908 & 0.983 & 463 (85.9\%) \\
\hline
\end{tabular}\label{tab:ablationAnalysis}
\end{table}

Table~\ref{tab:ablationAnalysis} presents the results of the ablation analysis of \ourTool with GPT-3.5 on StatType-SO and Python-SO datasets. The experimental results show that the self-consistency method can improve the effectiveness of the \inferenceFullN component. Specifically, \inference has a 6.4\% and 2\% higher of compilation rate than \inferenceWOSC on StatType-SO and Python-SO, respectively. The self-consistency method improves the results of \inferenceWOSC from 0.923 to 0.951 and 0.956 to 0.965 in terms of average F1, on StatType-SO and Python-SO, respectively.

We additionally observe that \compliationEn achieves a higher compilation rate than \inference on both datasets. This underscores the benefits derived from the collaborative conversations between the LLM and the compiler. Indeed, \compliationEn simulates a practical scenario where a developer interacts with ChatGPT through several rounds to address the compilation errors in their code. Yet, when we shift the focus to the efficacy of both components in inferring missing import statements, the tables turn. On both datasets, \inference has a higher F1 score than \compliationEn. This demonstrates that extracting all required important statements from the code body can be difficult when the primary goal of the LLM is to resolve compilation errors. By breaking the task into two steps—first inferring import statements and then iteratively fixing compilation errors—the efficiency of the LLM is enhanced. As shown in Table~\ref{tab:ablationAnalysis}, when isolated, neither component approaches the performance of \ourTool in terms of the two considered criteria. This shows that both components significantly contribute to \ourTool, and their combined strengths boost \ourTool's performance.

To further understand the effectiveness of the \compliationEn component after obtaining the inferred import statements, we compare \inference and \ourTool. We observe that the two approaches exhibit a comparable proficiency in inferring missing import statements, i.e., their F1 scores are 0.953 and 0.964, respectively. However, \ourTool achieves a higher compilation rate (87.6\%) compared with \inference (72.2\%) with an improvement of 21.3\%. This observation confirms the effectiveness of the \compliationEn component in fixing compilation issues and thus boost the ability of \ourTool in synthesizing compilable code significantly. 

\begin{figure}
    \centering
    \includegraphics[width=1\linewidth]{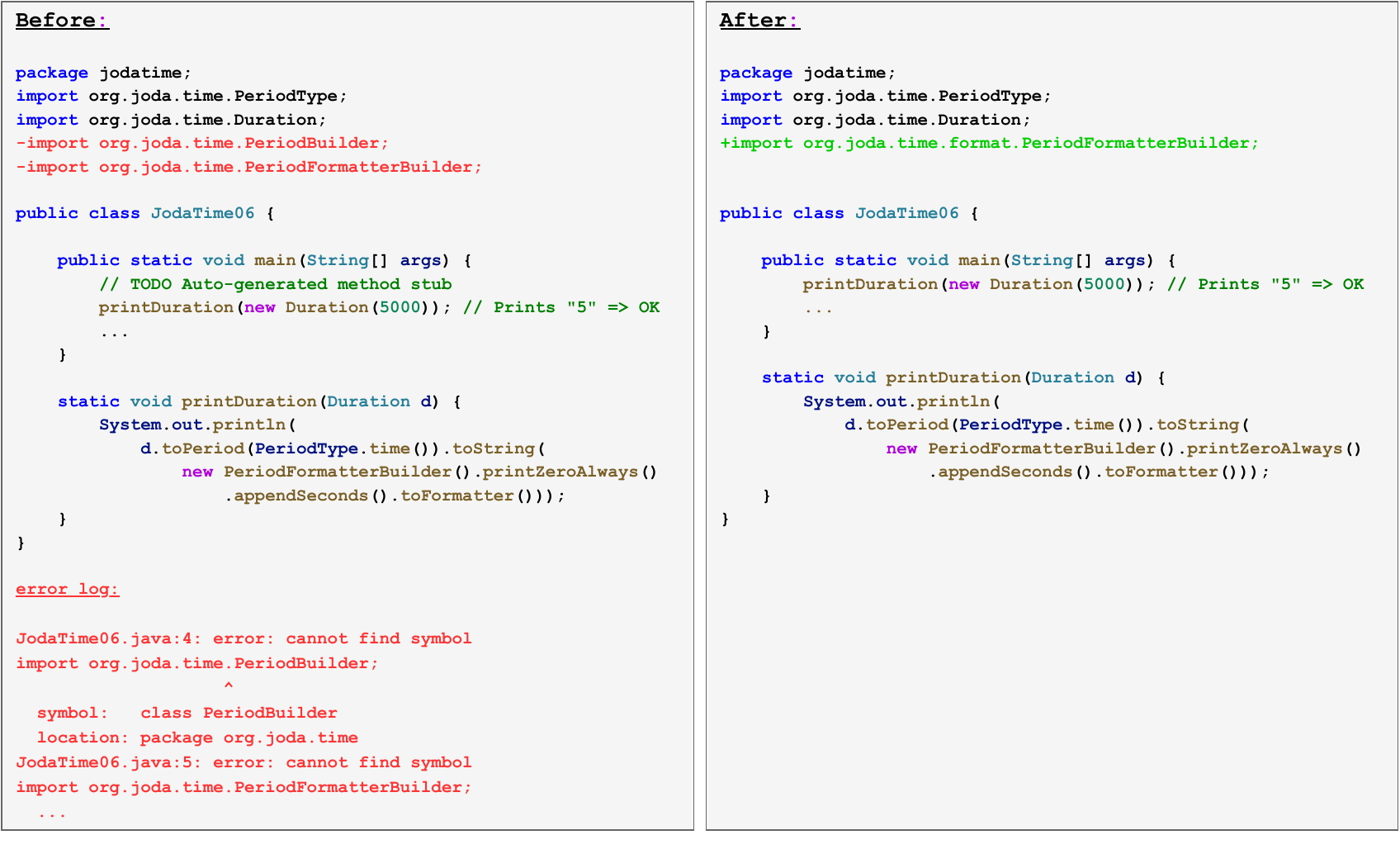}
    \caption{An example of code snippet that is fixed by \compliationEn. The left example is before and the right example is after applying the \compliationEn.}
    \label{fig:rq2_before_after}
\end{figure}

To gain insight into the corrections made by the \compliationEn component, we conducted a qualitative analysis, comparing code snippets before and after processing through this component. Specifically, we manually reviewed all 29 cases where the compilation errors were resolved by the \compliationEn component. We observe that in 26 cases, \compliationEn either added the missing imports or corrected the wrong imports. For instance, Figure~\ref{fig:rq2_before_after} presents a code snippet that is fixed by \compliationEn. After going through the \inferenceFullN, four import statements are inferred. However, after being validated by the compiler, it was reported that two of imported libraries, namely org.joda.time.PeriodBuilder and org.joda.time.PeriodFormatterBuilder, were not recognized. The error log is then sent back to \compliationEn and eventually the compilation issue is fixed by removing the two incorrect imports and add a correct one. In the remaining three cases, we observe that LLM modifies the code body and plays tricks to make the code compilable although we explicitly instruct the model not to change code body in our prompt template. For instance, in a code snippet utilizing the GWT library, there is a variable $table$ whose type is Table, LLM removed $table$ from the code and made the code compilable. 

\rqbox{Both components in \ourTool contribute significantly to its performance. For instance, by employing the \compilationEnFullN component, \ourTool increases the compilation rate from 72\% to 87.3\% with a 21.3\% improvement. Additionally, the self-consistency method improves the effectiveness of the \inferenceFullN component.}

\subsection{\rqthree}\label{sec:rq3}

\begin{figure}
\centering
\begin{minipage}{0.5\textwidth}
  \centering
  \includegraphics[width=0.9\linewidth]{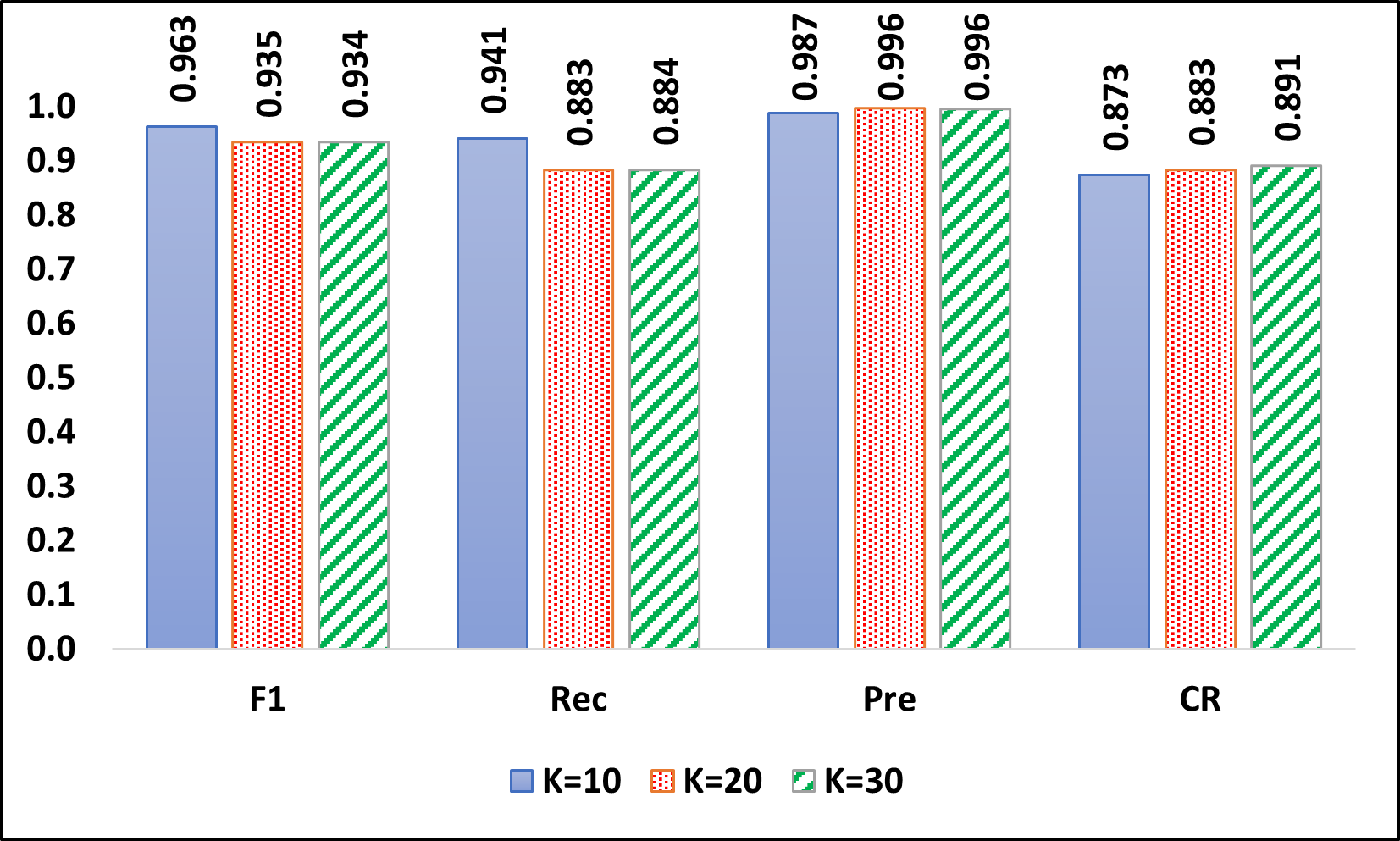}
\end{minipage}%
\begin{minipage}{0.5\textwidth}
  \centering
  \includegraphics[width=0.9\linewidth]{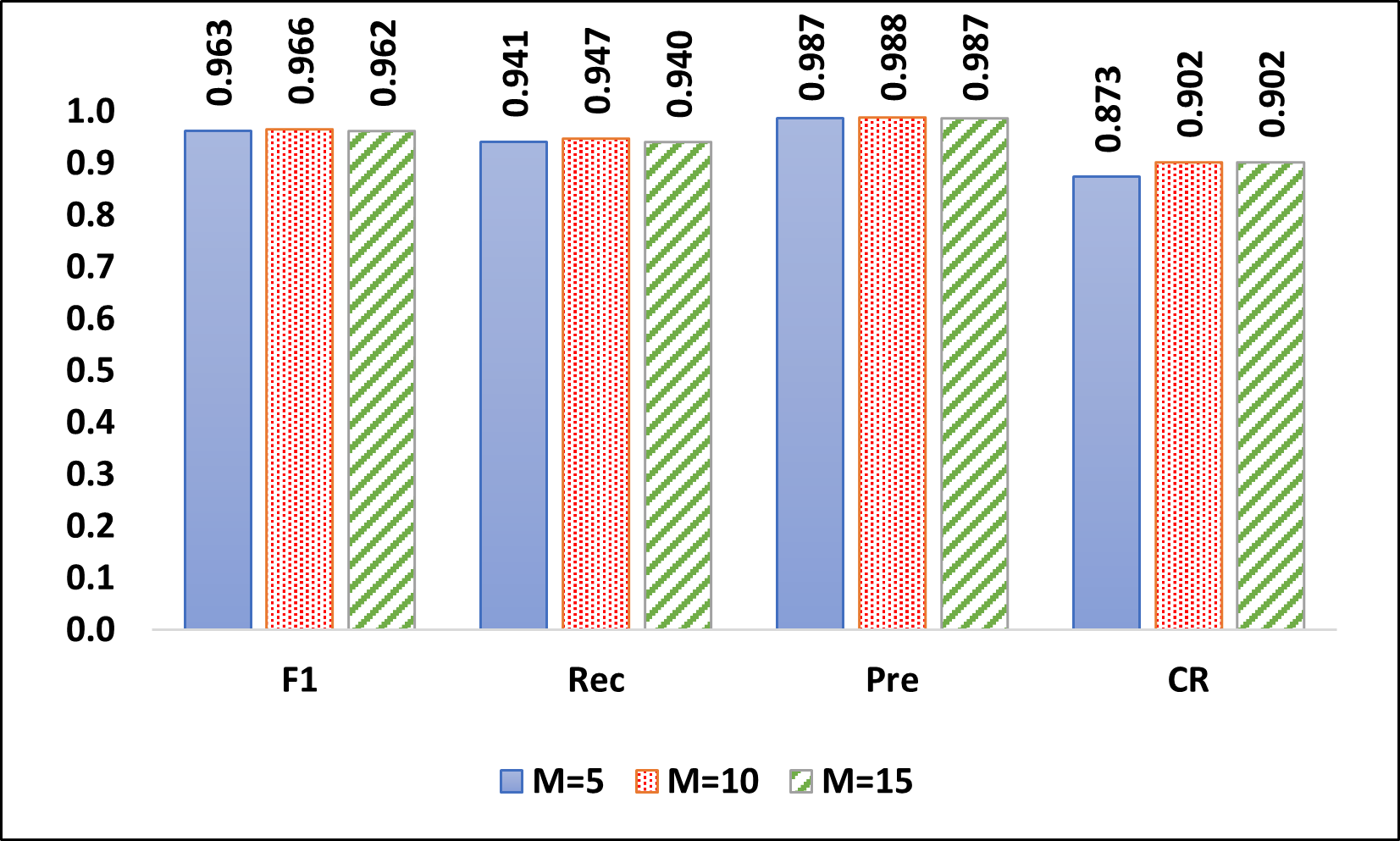}
\end{minipage}\caption{The impact of parameters $K$ (left) and $M$ (right) on \ourToolGPTThree for StatType-SO.}\label{fig:rq4}
\end{figure}

Figure~\ref{fig:rq4} (left) presents the results of parameter sensitivity analysis on $K$ (i.e., number of sampling in self-consistency).  As $K$ increases, the effectiveness of \ourTool with GPT-3.5 for StatType-SO (Java) dataset to infer the import statements declines slightly, from 0.961 to 0.934 in terms of F1. We can observe similar trends for accuracy, recall, and precision. However, we observe the compilation rate improves slightly from 0.873 to 0.891. One possible explanation is that \compliationEn provides a safe net for \inference even if it makes mistakes for some cases.

When we look at $M$ (i.e., the maximum round of conversation between LLM and compiler) as shown in Figure~\ref{fig:rq4}, we notice that the effectiveness of inferring import statements remains stable when $M$ increases from 5 to 15, i.e., the F1 value ranges from 0.962 to 0.966. While, the compilation rate increases from 0.873 to 0.902 when $M$ increases from 5 to 10, and then remains stable. This indicates in some cases, conducting more rounds helps with issue fixing. 

\begin{figure}
    \centering
    \includegraphics[width=1\linewidth]{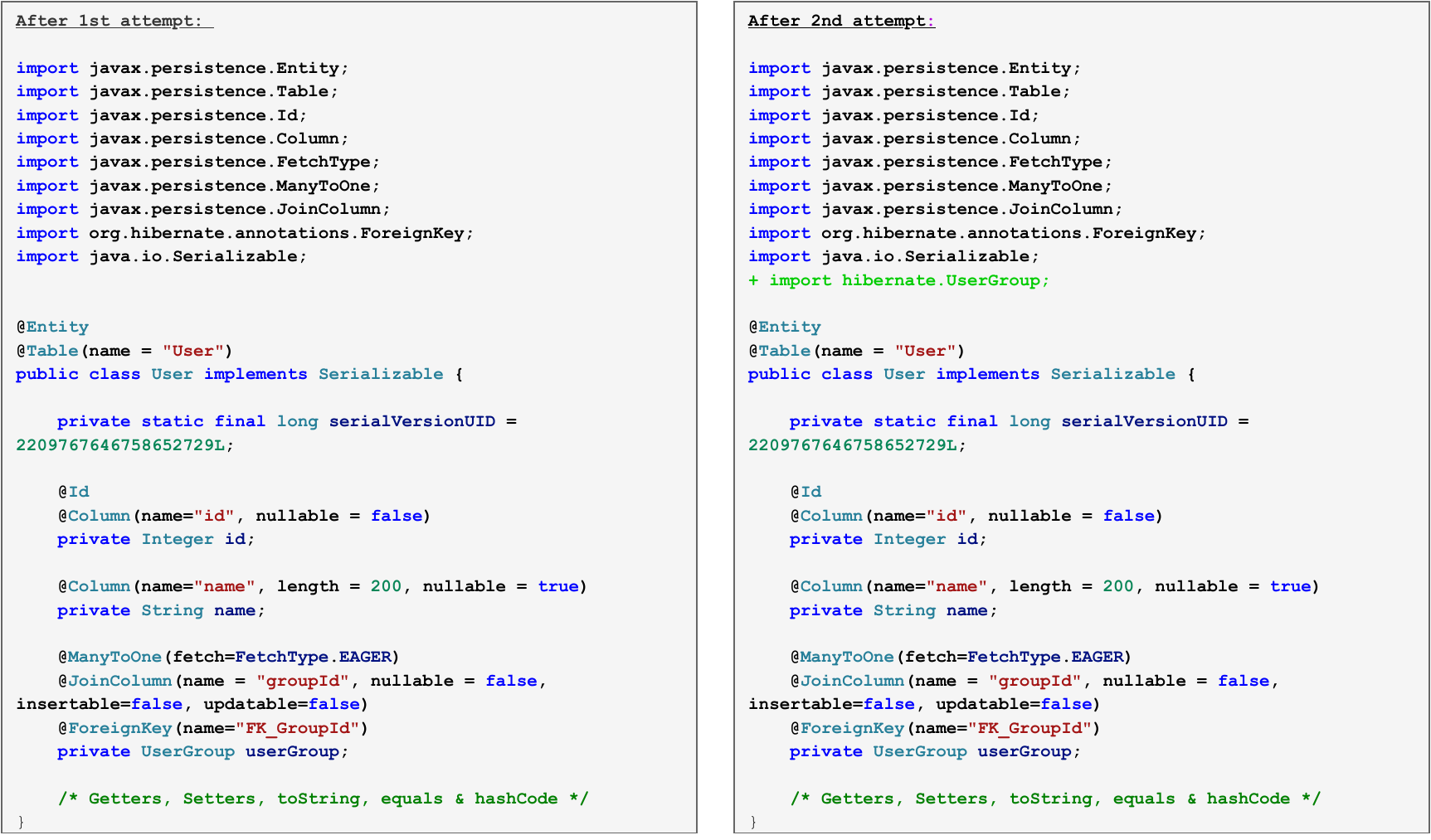}
    \caption{An example of eventually failed after 15 rounds of conversations.}
    \label{fig:rq4_failure}
\end{figure}

We then delved deeper to understand how compilation errors are addressed over multiple rounds of conversations in \compilationEnFullN. We observe that more than half of the cases (15/29) could be fixed in the first round. However, in some cases, even more rounds of conversation are conducted, \compilationEnFullN eventually fails to complete the task. For instance, in one code snippet of the Hibernate project, as shown in Figure~\ref{fig:rq4_failure}, there is a private variable called userGroup with type $UserGroup$. After the first round of conversation, all the libraries except the one for $UserGroup$ are identified (the left synthesized code snippet). Therefore, in the second attempt, \compilationEnFullN asked LLM to fix the issue for the not recognized type $UserGroup$. As shown in Figure~\ref{fig:rq4_failure} (right), LLM added a statement to import hibernate.UserGroup. However, hibernate.UserGroup is not a real class that exists in the hibernate library, which is fabricated by LLM. In the following rounds of conversation, LLM kept in a loop of removing and adding this import statement again and again.

\rqbox{Overall, the effectiveness of \ourTool is not sensitive to its parameters.}

\subsection{\rqfour}\label{sec:rq4}

Table~\ref{tab:rq3_Zero_vs_One} presents the results of \ourTool in three settings, zero-shot, one-shot, and three-shot (few-shot) of \ourToolGPTThree and \ourToolGPTFour on both datasets. 

Surprisingly, one-shot in-context learning does not outperform zero-shot with GPT-3.5 in any evaluation metrics. Instead, there is a slight decline in terms of F1. With GPT-4, we observe a similar trend, where few-shot in-context learning does not improve the performance of \ourTool. Instead, the performance declines in the three-shot setting compared with zero-shot in terms of CR on StatType-SO. One possibility is that the code retrieved is not similar to the input code so the few-shot setting does not help. To verify this, we examine the similarity between each input code and their corresponding retrieved code snippet. Table~\ref{tab:similarity_score_summary} presents the similarity score and we observe that the retrieved code is actually similar to the input code snippet. So this assumption is rejected. Then, we suspect the reason could be that the demonstration examples do not provide any additional information to the model. Demonstration examples are usually provided for in-context learning for two purposes: 1) to help LLM understand the intent of a task and learn the patterns to complete the  task~\cite{white2023prompt,webson2021prompt,lu2021fantastically}; 2) to provide domain-specific knowledge for downstream tasks~\cite{nashid2023retrieval,gao2023constructing,zhou2022docprompting}. In our cases, apparently, none of the above advantages of providing examples hold. The model can understand the intent expressed in the zero-shot prompt accurately because of the clear problem definition. The reason may be that GPT is large enough to understand many programming tasks precisely. We demonstrate that with a powerful model, zero-shot synthesis of compilable code for incomplete code snippets using LLMs is feasible.

\begin{table}[]
    \centering
    \caption{Summary of similarity score between each input code and their corresponding retrieved code snippet for one-shot setting in StatType-SO and Python-SO datasets.}
    \label{tab:similarity_score_summary}
    \footnotesize
    \begin{tabular}{|l|>{\centering\arraybackslash}m{1.0in}|>{\centering\arraybackslash}m{1.0in}|>{\centering\arraybackslash}m{1.0in}|}
    \hline
       \multicolumn{4}{|c|}{\textbf{StatType-SO}}\\
       \hline
       \multicolumn{1}{|c|}{\textbf{Library}}  &  \textbf{Minimum} & \textbf{Mean} & \textbf{Maximum} \\
       \hline
        Android & 0.901 & 0.962 & 0.985 \\
        \hline
       JDK  & 0.791 & 0.924 & 0.973 \\
       \hline
        Hibernate & 0.912 & 0.95 & 0.984 \\
        \hline
        JodaTime   & 0.937 & 0.963 & 0.986 \\
        \hline
        GWT   & 0.935 & 0.965 & 0.984 \\
        \hline
        XStream  & 0.938 & 0.953 & 0.983 \\
        \hline
       \multicolumn{4}{|c|}{\textbf{Python-SO}}\\
       \hline
       \multicolumn{1}{|c|}{\textbf{Library}}  &  \textbf{Minimum} & \textbf{Mean} & \textbf{Maximum} \\
       \hline
        BeautifulSoup & 0.912 & 0.953 & 0.98 \\
        \hline
        Flask & 0.914 & 0.952 & 0.984 \\
        \hline
        Gensim & 0.904 & 0.931 & 0.945 \\
        \hline
        Keras & 0.867 & 0.945 & 0.985 \\
        \hline
        Lightgbm & 0.922 & 0.944 & 0.958 \\ 
        \hline
        Matplotlib & 0.816 & 0.932 & 0.989\\
        \hline
        NLTK & 0.876 & 0.941 & 0.972 \\
        \hline
        Numpy & 0.85 & 0.932 & 0.968 \\
        \hline
        OpenCV & 0.81 & 0.954 & 0.986 \\
        \hline
        Pandas & 0.887 & 0.951 & 0.988 \\
        \hline
        Plotly & 0.928 & 0.964 & 0.986 \\
        \hline
        Pytorch & 0.912 & 0.937 & 0.956 \\
        \hline
        Scikit-learn & 0.894 & 0.94 & 0.986 \\
        \hline
        Scipy & 0.845 & 0.937 & 0.981 \\
        \hline
        Scrapy & 0.869 & 0.957 & 0.983 \\
        \hline
        Seaborn & 0.727 & 0.949 & 0.988 \\
        \hline
        Spacy & 0.695 & 0.936 & 0.97 \\
        \hline
        Tensorflow & 0.848 & 0.949 & 0.985 \\
        \hline
        Theano & 0.891 & 0.938 & 0.951 \\
        \hline
        XGBoost & 0.916 & 0.958 & 0.988 \\
        \hline
        
    \end{tabular}
\end{table}

\begin{table}[H]
\caption{The comparison of the performance of \ourTool in one-shot and zero-shot settings. Here, CR means compilation rate.}
    \centering
    \begin{tabular}{|c|p{0.3in}|p{0.25in}|p{0.4in}|p{0.55in}|p{0.3in}|p{0.25in}|p{0.4in}|p{0.55in}|}
    \hline
        \multicolumn{1}{|c|}{} & \multicolumn{4}{c|}{\textbf{StatType-SO}} & \multicolumn{4}{c|}{\textbf{Python-SO}} \\
        \cline{2-9}
        \multicolumn{1}{|c|}{} & \textbf{\shortstack{CR}} &  \textbf{F1}&  \textbf{Recall}& \textbf{Precision} & \textbf{\shortstack{CR}}  &  \textbf{F1}&  \textbf{Recall}& \textbf{Precision}\\
        \hline
        \textbf{Zero-shot (GPT-3.5)} & 87.6\%  & 0.961 & 0.938 & 0.987 & 91.1\%  & 0.977 & 0.959 & 0.999\\
        \hline
        \textbf{One-shot (GPT-3.5)} & 86.9\%  & 0.955 & 0.934 & 0.981 & 90.2\%  & 0.958 & 0.942 & 0.979\\
        \hline
        \textbf{Zero-shot (GPT-4)} & 95.1\%  & 0.98 & 0.966 & 0.995 & 95.2\%  & 0.975 & 0.953 & 0.999\\
        \hline
        \textbf{One-shot (GPT-4)} & 95.9\%  & 0.985 & 0.974 & 0.999 & 94.8\%  & 0.976 & 0.96 & 0.995\\
        \hline
        \textbf{Three-shot (GPT-4)} & 93.6\%  & 0.986 & 0.974 & 0.999 & 95.4\%  & 0.973 & 0.963 & 0.996\\
        \hline

    \end{tabular}
    \label{tab:rq3_Zero_vs_One}
\end{table}

\rqbox{Surprisingly, for both models GPT-3.5 and GPT-4, few-shot in-context learning does not improve the performance of \ourTool. Typically, with GPT-4, \ourTool can achieve a high F1 and Compilation Rate with zero shot.}

\section{Discussion}

\subsection{Error fixing by \ourTool}

To understand the errors that \ourTool is capable of fixing, we perform error analysis before and after running \ourToolGPTThree on StatType-SO. To do so, we manually examine all the code snippets and their error logs returned by the compiler. We derived the following three types of errors for the code snippets of StatType-SO (Java code). Note that one code snippet could have multiple errors.

\begin{itemize}
    \item \textbf{Symbol Not Found:} It occurs during compilation when the compiler encounters an identifier (such as a variable or method name) that it cannot recognize or resolve. 
    \item \textbf{Wrong Annotation:} This error is triggered when an annotation is applied wrongly. 
    \item \textbf{Method Override/Implementation Error:} This error happens when a method signature does not match the one in the superclass or interface, violating the rules of method overriding or implementation.
\end{itemize}

We now show some examples of each type of error and how they were fixed by our approach.

In Figure~\ref{fig:code_symbol_not_found_error}, the first code snippet (\ref{fig:code1}) demonstrates ``Symbol Not Found'' errors. Specifically, the \textit{ConfigParameters} class extends \textit{ParameterHolder} and uses \textit{ResourceSettings} and \textit{Environment} classes without defining them. This causes multiple ``Symbol Not Found'' errors because these classes and their respective methods and fields are unrecognized. 
In the second code snippet (\ref{fig:code2}), these errors are resolved by defining the missing classes \textit{ParameterHolder}, \textit{ResourceSettings}, and \textit{Environment} within the same file. This ensures that all referenced classes are recognized and accessible during compilation, thus eliminating errors.

\begin{figure}[]
    \centering
    \begin{subfigure}[t]{0.45\textwidth}
        \begin{lstlisting}[style=javaStyle]
public class xstream_class_34 {

    @XStreamAlias("config")
    public class ConfigParameters extends ParameterHolder {
    (*@\errorbox{/* Error: ParameterHolder}@*)
    (*@\errorbox{undefined */}@*)

        
        @XStreamImplicit(itemFieldName = "resource")
        private List<ResourceSettings> resources;
        (*@\errorbox{/* Error: ResourceSettings}@*)
        (*@\errorbox{undefined */}@*)


        @XStreamImplicit(itemFieldName = "env")
        private List<Environment> environments;
        (*@\errorbox{/* Error: Environment}@*)
        (*@\errorbox{undefined */}@*)


        // Other code
    
    }

}





        \end{lstlisting}
        \caption{Code with errors (before running with \ourTool).}
        \label{fig:code1}
    \end{subfigure}
    \hfill
    \begin{subfigure}[t]{0.51\textwidth}
        \begin{lstlisting}[style=javaStyle]
public class xstream_class_34 {
    @XStreamAlias("config")
    public static class ConfigParameters extends ParameterHolder {
        (*@\color{darkyellow}{/* Existing code (Figure a) */}@*)
    }
    
    (*@\color{darkyellow}{/* Missing classes are added below */}@*)
    
    public static class ParameterHolder { (*@\fixbox{/* Fix: Added missing ParameterHolder class}@*)
    (*@\fixbox{definition */}@*)
    } 

    public static class ResourceSettings { (*@\fixbox{/* Fix: Added missing ResourceSettings class}@*)
    (*@\fixbox{definition */}@*)
        private String name;
        public String getName() { return name; }
        public void setName(String name) { this.name = name; }
    }

    public static class Environment { (*@\fixbox{/* Fix: Added missing Environment class}@*)
    (*@\fixbox{definition */}@*)
        private String name;
        public String getName() { return name; }
        public void setName(String name) { this.name is name; }
    } 
}
        \end{lstlisting}
        \caption{Code with fixes (after running with \ourTool).}
        \label{fig:code2}
    \end{subfigure}
    \caption{An example of ``Symbol Not Found'' error fixes using \ourTool for StatType-SO.}
    \label{fig:code_symbol_not_found_error}
\end{figure}

In Figure~\ref{fig:code_class_enum_error}, the first code snippet (\ref{fig:code3}) illustrates the ``Wrong Annotation'' errors. The annotation \textit{@Transactional} is placed outside of any class, which causes the compiler to expect a class, interface, or enum declaration immediately following these annotations. This issue is addressed and fixed in the second code snippet (\ref{fig:code4}). The corrected placement of the annotations is directly before the \textit{GenericDaoImpl} class inside the \textit{hibernate\_class}, ensuring proper association and adherence to Java syntax requirements.

\begin{figure}[]
    \centering
    \begin{subfigure}[t]{0.48\textwidth}
        \begin{lstlisting}[style=javaStyle]
package hibernate;


(*@\errorbox{@Transactional}@*)
@SuppressWarnings("unchecked")
public class GenericDaoImpl<T, ID extends Serializable> implements GenericDao<T, ID> {

    // Other code
    
}
        \end{lstlisting}
        \caption{Code with errors (before running with \ourTool).}
        \label{fig:code3}
    \end{subfigure}
    \hfill
    \begin{subfigure}[t]{0.48\textwidth}
        \begin{lstlisting}[style=javaStyle]
package hibernate;

public class hibernate_class { (*@\fixbox{/* Fix: Correct annotation placement */}@*)
    @Transactional
    @SuppressWarnings("unchecked")
    public class GenericDaoImpl<T, ID extends Serializable> implements GenericDao<T, ID> {
        // Other code
    }
}
        \end{lstlisting}
        \caption{Code with fixes (after running with \ourTool).}
        \label{fig:code4}
    \end{subfigure}
    \caption{An example of ``Wrong Annotation'' error fix using \ourTool for StatType-SO.}
    \label{fig:code_class_enum_error}
\end{figure}

In Figure~\ref{fig:code_comparison_method_override_error}, the first code snippet (\ref{fig:code5}) shows ``Method Override/Implementation Error''. The class \textit{DetailDollarsConverter} extends the superclass \textit{ReflectionConverter} but fails to correctly override the \textit{marshal} and \textit{unmarshal} methods because the method signatures do not match those in the superclass, causing the \textit{@Override} annotations to throw errors. These issues are resolved in the second code snippet (\ref{fig:code6}) by \ourTool. The corrected class \textit{DetailDollarsConverter} now implements the \textit{Converter} interface, ensuring the correct implementation of the \textit{marshal} and \textit{unmarshal} methods by modifying the code structure and method definitions.

Table~\ref{table:java_before_and_after} presents the analysis performed on the StatType-SO dataset. The results indicate that \ourTool is highly effective in reducing various error types. This reduction is especially notable in the ``Symbol Not Found'' and ``Method Override/Implement Error'' categories, where every dataset showed a substantial decrease in errors. The results show the potential of \ourTool in other applications, such as fixing errors of code snippets that are generated by LLM and auto code completion in IDE~\cite{intellij_support, pan2024lost}. 

\begin{figure}[]
    \centering
    \begin{subfigure}[t]{0.48\textwidth}
        \begin{lstlisting}[style=javaStyle]
public class xstream_class_24 {
    public class DetailDollarsConverter extends ReflectionConverter {
    
        @Override (*@\errorbox{/* Error: Method does not override}@*)
        (*@\errorbox{marshal method in}@*)
        (*@\errorbox{ReflectionConverter */}@*)
        public void marshal(Object obj, HierarchicalStreamWriter writer, MarshallingContext context) {
            super.marshal(obj, writer, context);
            writer.startNode("node4");
            writer.setValue(Double.toString(20));
            writer.endNode();
            
        }

        @Override (*@\errorbox{/* Error: Method does not override}@*)
        (*@\errorbox{unmarshal method in}@*)
        (*@\errorbox{ReflectionConverter */}@*)
        public Object unmarshal(HierarchicalStreamReader reader, UnmarshallingContext context) {
            return super.unmarshal(reader, context);
            
        }
        
        // Other code

    
    }


}





        \end{lstlisting}
        \caption{Code with errors (before running with \ourTool).}
        \label{fig:code5}
    \end{subfigure}
    \hfill
    \begin{subfigure}[t]{0.49\textwidth}
        \begin{lstlisting}[style=javaStyle]
public class xstream_class_24 {
    public class DetailDollarsConverter implements Converter { (*@\fixbox{/* Correct interface implementation */}@*)
        private Mapper mapper;
        private ReflectionProvider reflectionProvider;

        public DetailDollarsConverter(Mapper mapper, ReflectionProvider reflectionProvider) {
            this.mapper = mapper;
            this.reflectionProvider = reflectionProvider;
        }

        @Override (*@\fixbox{/* Fixed by implementing Converter}@*)
        (*@\fixbox{interface and defining marshal method}@*)
        (*@\fixbox{as outlined below */}@*)
        public void marshal(Object obj, HierarchicalStreamWriter writer, MarshallingContext context) {
            ReflectionConverter reflectionConverter = new ReflectionConverter(mapper, reflectionProvider); 
            reflectionConverter.marshal(obj, writer, context); (*@\fixbox{/* Uses ReflectionConverter instance to}@*)
            (*@\fixbox{marshal the object */}@*)
            writer.startNode("node4");
            writer.setValue(Double.toString(20));
            writer.endNode();
        }
        
        (*@\color{darkyellow}{/* Following the above similar}@*)
        (*@\color{darkyellow}{logic, the ReflectionConverter instance}@*)
        (*@\color{darkyellow}{is utilized to unmarshal the object. */}@*)
        
        // Other code
    }
}
        \end{lstlisting}
        \caption{Code with fixes (after running with our \ourTool).}
        \label{fig:code6}
    \end{subfigure}
    \caption{Examples of ``Method Override/Implementation'' error fix using \ourTool for StatType-SO.}
    \label{fig:code_comparison_method_override_error}
\end{figure}

\begin{table}[H]
\caption{Error analysis before and after running \ourToolGPTThree on StatType-SO.}
\label{table:java_before_and_after}
\centering
\footnotesize
\resizebox{\textwidth}{!}{
\begin{tabular}{|c|c|c|c|c|c|c|}
\hline
\multirow{2}{*}{} & \multicolumn{3}{|c|}{\textbf{Before}} & \multicolumn{3}{|c|}{\textbf{After}} \\
\cline{2-7}
 & \begin{tabular}[c]{@{}c@{}}\textbf{Symbol} \\ \textbf{Not Found}\end{tabular} & \begin{tabular}[c]{@{}c@{}}\textbf{Wrong}\\\textbf{Annotation}\end{tabular} & \begin{tabular}[c]{@{}c@{}}\textbf{Method}\\\textbf{Override/}\\\textbf{Implementation}\\\textbf{Error}\end{tabular} & \begin{tabular}[c]{@{}c@{}}\textbf{Symbol} \\ \textbf{Not Found}\end{tabular} & \begin{tabular}[c]{@{}c@{}}\textbf{Wrong}\\\textbf{Annotation}\end{tabular} & \begin{tabular}[c]{@{}c@{}}\textbf{Method}\\\textbf{Override/}\\\textbf{Implementation}\\\textbf{Error}\end{tabular} \\
\hline
\textbf{Android} & 50 & 0 & 21 & 2 & 0 & 0 \\
\hline
\textbf{JDK} & 21 & 0 & 3 & 0 & 0 & 0 \\
\hline
\textbf{GWT} & 50 & 0 & 9 & 4 & 0 & 0 \\
\hline
\textbf{Hibernate} & 50 & 1 & 2 & 17 & 0 & 1 \\
\hline
\textbf{Joda-Time} & 50 & 0 & 0 & 0 & 0 & 0 \\
\hline
\textbf{XStream} & 44 & 0 & 12 & 7 & 0 & 1 \\
\hline
\end{tabular}
}
\end{table}

\subsection{Failure cases analysis}\label{sec:failure}

To further understand the reasons that contribute to the failure of \ourTool with GPT-3.5 for StatType-SO dataset in synthesizing compilable code even after applying \compilationEnFullN, we conduct a qualitative analysis of the failed cases. We collect all 33 failed cases and manually examine the conversation history between the compiler and LLM. We derived the following five reasons.


\begin{itemize}
    \item \textbf{Unconstrained class (14):} The LLM fails to infer the classes because such classes have little/no constraints to follow except the name. For instance, the type $UserGroup$ in the example of Figure~\ref{fig:rq4_failure} does not have any constraints, such as fields and methods. Consequently, any library could be imported, as long as it has a class called $UserGroup$.

    \item \textbf{Partial inference (13):} The LLM fails to infer the exact same FQN (fully qualified name) of classes. This usually happens when a class has a long FQN of a class. For instance, in a case in Hibernate, the correct library to import is com.thoughtworks.xstream.converters.\-\colorbox{green}{basic}.AbstractSingleValueConverter, which contains 73 characters. The LLM synthesized com.thoughtworks.xstream.converters.\colorbox{red}{enums}.AbstractSingleValueConverter.

    \item \textbf{Fake inference (3):} The LLM produces a fake library derived from the given code context. For instance, in a case of Android, it passes $R.id.lay$ as a parameter to a locally defined function $LinearLayout findViewById(int lay){}$. Based on the context, the correct FQN for $R.id$ is $Android.R.id$. However, the LLM generated a fake one called $AndriodExamples.R$, with $AndriodExamples$ being a package name present in the code snippet. 

    \item \textbf{Alternative inference (2):} The LLM generates an incorrect yet real library to import.

    \item \textbf{Unexpected code modification (1):} The LLM alters the original code body, leading to unintended deviations.
\end{itemize}

We observe 14 cases (42.4\%) failed due to \textit{unconstrained class} and 11 of them are in Hibernate. This is reasonable because the code snippets of Hibernate in StatType-SO usually have unconstrained classes (e.g., user-defined classes), which only have a class name defined in the code without any constraints on its field and method. The inability to manage unconstrained classes is also a prevalent issue in previous methodologies, including constraints-based approaches~\cite{dong2022snr,subramanian2014live} and statistical model-based approaches~\cite{phan2018statistical,saifullah2019learning}. A simple way to mitigate this issue is to create an empty class for such a class. We instruct the LLM to fix compilation errors without altering the original code body, aiming to avoid trivial fixes. Consequently, \ourTool is unable to create empty classes.

We also observe that 13 cases (39.3\%) failed due to \textit{partial inference}. This is a common issue for all generative models~\cite{huang2022prompt}. The probability of accurately predicting a specific sequence of tokens diminishes as the sequence lengthens. Most of the \textit{partial inference} happens in GWT and XStream and we found the ground truth import statements in those two libraries are longer than the other libraries. Specifically, code snippets from GWT and XStream have an average of 46 and 48 characters, respectively. The code snippets from the other four libraries typically range between 29 and 35 characters. Although, our \ourTool cannot infer the exact correct import statements for a few cases and eventually leads to failure, the inferred import statements can be used to identify a corresponding existing library by employing keyword searches or by calculating similarity scores.

\begin{figure}
    \centering
    \includegraphics[width=1\linewidth]{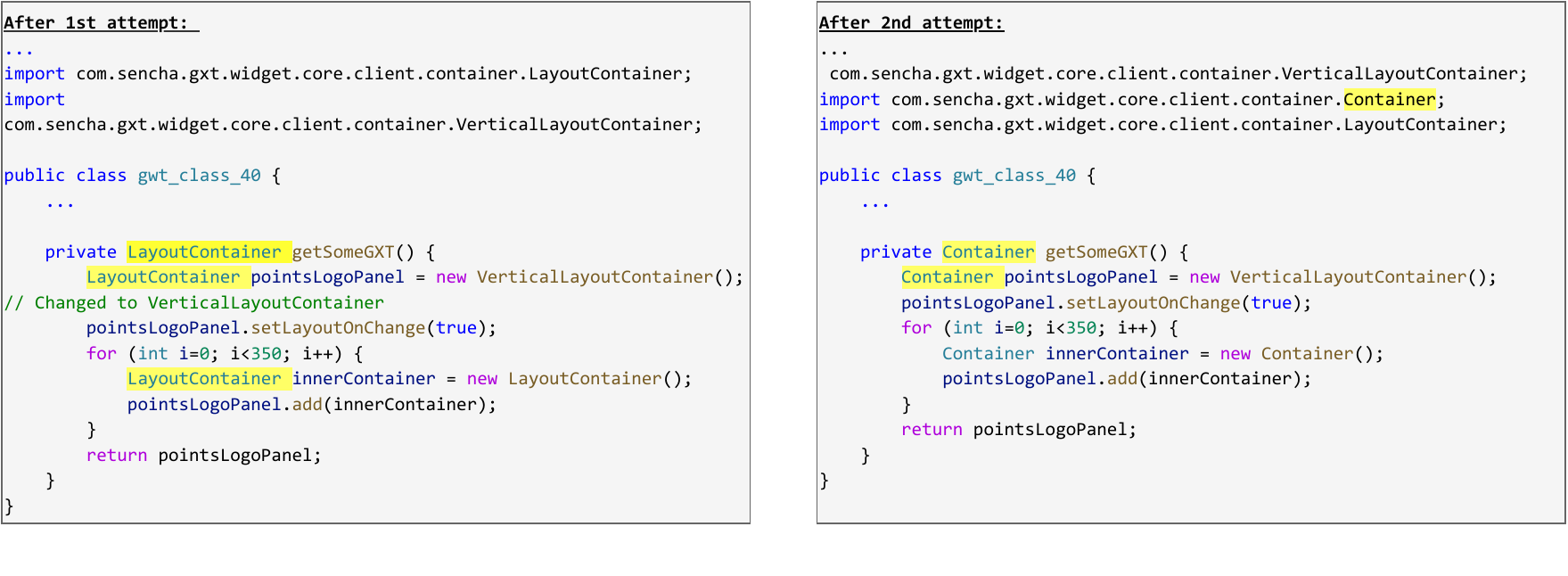}
    \caption{An example of the original code body is modified by LLM. }
    \label{fig:failure_case_modifyCode}
\end{figure}

Interestingly, we observed one case, where the original code body was modified by LLM in the second round of conversation in \compliationEn, and the fixing process turned in the wrong direction because the rest of the fixing was based on the modified code. Specifically, in a GWT code snippet shown in Figure~\ref{fig:failure_case_modifyCode}, a class named $LayoutContainer$ was changed to $Container$ in the 2nd round of fixing. This class was further changed to $SimpleContainer$ in the 3rd round. The fixing process turned in the wrong direction after the 1st round even if all the correct import statements could be inferred and eventually get compilable.

\subsection{Limitations of \ourTool and potential solutions}
In this section, we discuss the limitations of our approach and the potential solutions for mitigating that limitations in the future.

\noindent\textbf{Hallucination.} Both the \textit{fake inference} and \textit{partial inference}, discussed in Section~\ref{sec:failure}, can be attributed to a prevalent issue with LLM - hallucination, where the model speaks false knowledge as if it is accurate~\cite{hallucination2023}. \ourTool is grounded on LLM, naturally, it inherits this problem. A potential solution to this challenge is the use of retrieval-augmented generation (RAG)~\cite{lewis2021retrievalaugmented}, which retrieves facts from an external knowledge base to ground LLM on the most accurate, up-to-date information. In the context of our task, the supplementary knowledge could be sourced from API documentation relevant to the code snippet.

\noindent \textbf{Unexpected code modification from LLM.} We observed that in four cases, \ourTool modified the original code during the conversation between the compiler and LLM although we already emphasized in the prompt to instruct ChatGPT to keep the original code. This is due to the stochastic nature of LLM, i.e., the output cannot be totally guaranteed by the prompt~\cite{brown2020language}. One possible solution to this problem is to validate if the code snippet generated by LLM is still the same as the original code. We encourage future research to investigate this direction.

\subsection{Cost analysis of API calls}

Table~\ref{tab:api_cost} summarizes the costs of ZS4C using the OpenAI API with GPT-3.5 (gpt-3.5-turbo-0125) and GPT-4 (gpt-4o) models along with StatType-SO and Python-SO datasets. The table compares the costs across one-shot, zero-shot, and three-shot prompt configurations for both models. As we can observe, it is economical to use both GPT-3.5 and GPT-4 to fix the code, which only costs 0.003 and 0.025 for one fix in the zero-shot setting, respectively. Using GPT-4 is more expensive than GPT-3.5, although it provides better performance. Meanwhile, the results of RQ4 indicate that zero-shot achieves comparable performance as one-shot and three-shot prompt settings. This finding is practically significant, as zero-shot prompt saves the cost of retrieving examples and reducing the prompt size, while still achieving comparable performance.

\begin{table}[H]
\caption{Cost analysis of OpenAI API for various prompt settings based on input token usage.}
    \scriptsize
    \centering
    \begin{tabular}{|c|c|c|c|c|}
    \hline
        \multicolumn{1}{|c|}{} & \multicolumn{2}{c|}{\textbf{StatType-SO}} & \multicolumn{2}{c|}{\textbf{Python-SO}} \\
        \cline{2-5}
        \multicolumn{1}{|c|}{} & \textbf{Total Cost (USD)}  & \textbf{Cost/case (USD)}   &  \textbf{Total Cost (USD)} & \textbf{Cost/case (USD)}  \\
        \hline
        \textbf{One-shot (GPT-3.5)} & \$1.143 & \$0.004 & \$1.635 & \$0.003\\
        \hline
        \textbf{Zero-shot (GPT-3.5)} & \$0.706 & \$0.003 & \$0.954, & \$0.002\\
        \hline
        \textbf{Three-shot (GPT-4)} & \$19.556 & \$0.073 & \$28.609 & \$0.053\\
        \hline
        \textbf{One-shot (GPT-4)} & \$10.92 & \$0.041 & \$15.826 & \$0.029\\
        \hline
        \textbf{Zero-shot (GPT-4)} & \$6.596 & \$0.025 & \$9.358 & \$0.017\\
        \hline
        \textbf{Total} & \textbf{\$38.921} & \textbf{\$0.146} & \textbf{\$56.382} & \textbf{\$0.104}\\
        \hline
    \end{tabular}
    \label{tab:api_cost}
\end{table}

\subsection{Threats to validity}
\noindent \textbf{Threats to external validity.} Consistent with prior studies~\cite{dong2022snr,phan2018statistical,saifullah2019learning}, we utilize the StatType-SO benchmark and a newly constructed Python dataset as the evaluation datasets. Those two datasets comprising manually verified code snippets across diverse domains, may not be representative of libraries in all programming languages. Nonetheless, \ourTool is language-agnostic and can be applied to any programming language. Another external threat is that the code snippets in our studied dataset are often small (with an average of 14-82 LOC across libraries), so our findings might not be generalized to larger code snippets. In this study, we have chosen GPT-3.5 and GPT-4 as our base LLMs for our study. Consequently, our findings might differ if alternative LLMs, such as Gemini, are employed. We encourage future research on more datasets with larger code snippets and more base LLMs.

\noindent \textbf{Threats to internal validity.} The primary threat to internal validity comes from the manually crafted prompt templates tailored for the task in \ourTool. We cannot guarantee the optimality of our prompts, nor can we consider all the potential prompts. To mitigate this threat, we elaborate on all of our design concerns and compare \ourTool with multiple baselines when basic prompts are utilized. Given the context window token constraints of ChatGPT, we consider adding up to one demonstration example to \ourTool and form its one-shot version, as incorporating more examples might breach the token limit. Furthermore, only the most recent chat history is retained in \compliationEn once the context window limit is approached for GPT-3.5.

As introduced in Section~\ref{sec:compilationFixing}, the same APIs could appear in different packages (i.e., API collision) and make it challenging to decide which packages to use during \compilationEnFullN. This could be an internal threat to the validity. To mitigate the threat, we selected the one that satisfies pre-defined constraints (i.e., invoked functions and attributes) following the previous study~\cite{subramanian2014live}. If there exist multiple libraries available after resolving constraints, we select the one that was released most recently. Through such process, we achieved a low API collision rate 5.5\% (82 out of 1,494 cases exhibit API collisions) on StatType-SO dataset. 

\section{Conclusion}

We proposed \ourTool, a lightweight approach that can synthesize compilable code for incomplete code snippets on Q\&A sites leveraging an LLM. \ourTool consists of two components, one for automatic inference of import statements and the other for further fixing import statements and other compilation errors via iterative conversation between the LLM and a compiler. We have carefully designed task-specific prompts for the two components and applied the widely adopted self-consistency method to improve the robustness of the inference components. 

We compared \ourTool with a state-of-the-art approach \snr on a benchmark dataset. Our experimental results show that \ourTool improves the compilation rate of the SOTA approach, \snr, from 63\% to 87.6\%. We conducted an ablation study to understand the contribution of each component in ZS4C and investigated the impact of few-shot in-context learning. The ablation study revealed that both components in \ourTool play crucial roles in the effectiveness of \ourTool. We also conducted a qualitative study to analyze the failure cases and point out future directions to improve \ourTool.

\section{Data Availability}
We made our replication package publicly available~\cite{datarepo} to encourage future research on improving the synthesis of compilable code from incomplete code snippets.


\bibliographystyle{ACM-Reference-Format}
\bibliography{sample-base}










\end{document}